\documentclass[aps,prl,superscriptaddress,showpacs,twocolumn]{revtex4-1}
\usepackage[left=1.9cm,right=1.9cm,bottom=1.9cm,top=1.9cm,ignoreall]{geometry}
\usepackage[utf8]{inputenc}
\usepackage{graphicx}
\usepackage{amsmath}
\usepackage{amssymb}
\usepackage{xspace}
\usepackage{bm}
\usepackage{color}
\usepackage[squaren,Gray]{SIunits}
\usepackage[hyperindex=true]{hyperref}
\hypersetup{linktocpage,colorlinks=true,citecolor=blue,linkcolor=blue}
\usepackage{empheq}
\usepackage{braket}
\usepackage{physics}

\begin{document}
	
	\title{Engineering two-photon wavefunction and exchange statistics\\ in a semiconductor chip}

	\author{S. Francesconi}
	\affiliation{Laboratoire Matériaux et Phénomènes Quantiques, Université de Paris, CNRS-UMR 7162, Paris 75013, France}
	\author{F. Baboux}
	\thanks{Corresponding author: florent.baboux@u-paris.fr}
	\affiliation{Laboratoire Matériaux et Phénomènes Quantiques, Université de Paris, CNRS-UMR 7162, Paris 75013, France}
	\author{A. Raymond}
	\affiliation{Laboratoire Matériaux et Phénomènes Quantiques, Université de Paris, CNRS-UMR 7162, Paris 75013, France}
	\author{N. Fabre}
	\affiliation{Laboratoire Matériaux et Phénomènes Quantiques, Université de Paris, CNRS-UMR 7162, Paris 75013, France}
	\author{G.~Boucher}
	\affiliation{Laboratoire Matériaux et Phénomènes Quantiques, Université de Paris, CNRS-UMR 7162, Paris 75013, France}
	\author{A. Lemaître}
	\affiliation{Université Paris-Saclay, CNRS, Centre de Nanosciences et de Nanotechnologies, 91120, Palaiseau, France}
	\author{P. Milman}
	\affiliation{Laboratoire Matériaux et Phénomènes Quantiques, Université de Paris, CNRS-UMR 7162, Paris 75013, France}
	\author{M. Amanti}
	\affiliation{Laboratoire Matériaux et Phénomènes Quantiques, Université de Paris, CNRS-UMR 7162, Paris 75013, France}
	\author{S. Ducci}
	\affiliation{Laboratoire Matériaux et Phénomènes Quantiques, Université de Paris, CNRS-UMR 7162, Paris 75013, France}
	
	\date{\today}

\begin{abstract}

High-dimensional entangled states of light provide novel possibilities for quantum information, from fundamental tests of quantum mechanics to enhanced computation and communication protocols. 
In this context, the frequency degree of freedom combines the assets of robustness to propagation and easy handling with standard telecommunication components.
Here we use an integrated semiconductor chip to engineer the wavefunction and exchange statistics of frequency-entangled photon pairs directly at the generation stage, without post-manipulation.
Tailoring the spatial properties of the pump beam allows to generate frequency-anticorrelated, correlated and separable states, and to control the symmetry of the spectral wavefunction to induce either bosonic or fermionic behaviors.
These results, obtained at room temperature and telecom wavelength, open promising perspectives for the quantum simulation of fermionic problems with photons on an integrated platform, as well as for communication and computation protocols exploiting antisymmetric high-dimensional quantum states.

\end{abstract}

\maketitle
	
\section{Introduction}

Nonclassical states of light are key resources for quantum information technologies thanks to their easy transmission, robustness to decoherence and variety of degrees of freedom to encode information \cite{Walmsley15}. In recent years, great efforts have been directed towards entanglement in high-dimensional degrees of freedom of photons as a means to strengthen the violation of Bell inequalities \cite{Collins02,Dada11}, increase the density and security of quantum communication \cite{Cerf02,Barreiro08} and enhance flexibility in quantum computing \cite{Lanyon09}. In addition, high-dimensional degrees of freedom of photons display a perfect analogy with the continuous variable (CV) of a multiphoton mode of the electromagnetic field \cite{Abouraddy07}, which make them a promising platform to realize CV quantum information protocols in the few-photon regime \cite{Tasca11,Fabre19}. Photonic high-dimensional entanglement has been recently demonstrated into orbital angular momentum \cite{Dada11,Fickler12}, transverse spatial \cite{Walborn03} and path \cite{Solntsev17,Wang18} modes, and frequency (or frequency-time) \cite{Kues17,Ansari18} degrees of freedom.

Among these different degrees of freedom, frequency is particularly attractive thanks to its robustness to propagation in optical fibers and its capability to convey large-scale quantum information into a single spatial mode. This provides a strong incentive for the development of efficient and scalable methods for the generation and the manipulation of frequency-encoded quantum states \cite{Ansari18b,MacLean18,Davis18}. Nonlinear parametric processes such as parametric down-conversion (PDC) and four-wave mixing offer a high versatility for the generation of frequency-entangled photon pairs \cite{Chen19,Kues19}. However, under CW pumping energy conservation naturally leads to the emission of frequency-anticorrelated states, whereas other types of correlations are needed for certain applications: for instance non-correlated states are required for heralded single photon sources \cite{Mosley08,Belhassen18} and correlated states are key resources for clock synchronization \cite{Giovannetti01} or dispersion cancellation in long-distance communication \cite{Lutz14}.
At a deeper level, it is desirable to gain a higher control over the frequency degree of freedom by manipulating the biphoton joint spectrum both in amplitude and phase. Such shaping can be performed by post-manipulation using time lenses \cite{Donohue16}, spatial light modulators (SLM) \cite{Peer05,Bernhard13}, dispersive elements \cite{Jin18} or programmable phase filters \cite{Kues17}, but this inevitably  reduces the brightness of the source and its integrability into chip-based photonic circuits.
Direct shaping of quantum frequency states at the generation stage is therefore preferable. Using parametric processes in solid-state systems, this has been recently realized by engineering the spectral \cite{Mosley08,Kumar14,Ansari18,Ansari18c} and spatial \cite{Valencia07} properties of the pump beam, by temperature tuning \cite{Tischler15} or by tailoring the material nonlinearity in domain-engineered crystals \cite{Graffitti18}.
Among these different approaches, the spatial tuning of the pump combines the advantages of reconfigurability and extended possibilities of frequency state engineering \cite{Boucher15}. However, to our knowledge no previous work  has demonstrated a complete toolbox for frequency state engineering through pump spatial tuning, including a control over the symmetry of the joint spectrum and thus the exchange statistics of the photon pairs -- an important feature of quantum state engineering though, in particular in view of quantum simulation \cite{Crespi13,Matthews13,Crespi15}.

In this work, we exploit the high flexibility offered by PDC in a semiconductor AlGaAs microcavity under a transverse pump geometry \cite{Walton03,Caillet09,Orieux13} to engineer the spectral wavefunction and exchange statistics of photon pairs without post-manipulation. Tuning the pump spatial intensity allows to produce frequency-anticorrelated, correlated and separable states, while tuning the spatial phase enables to switch between symmetric and antisymmetric spectral wavefunctions, leading respectively to bosonic and fermionic behaviors in a quantum interference experiment \cite{Walborn03,Fedrizzi09}. 
We also demonstrate the generation of non-Gaussian entanglement \cite{Gomes09,Douce13} in the continuous variables formed by the frequency and time degrees of freedom of the photon pairs. We thus demonstrate a general method providing a complete toolbox for frequency state engineering, at the generation stage and using a chip-based source: these characteristics are crucial in the perspective of the real-world deployment of photonic quantum technologies based on the frequency degree of freedom.
Our results, obtained at room temperature and telecom wavelength, open promising perspectives for quantum simulation with particles of various statistics on a monolithic platform without requiring external sources of quantum light \cite{Crespi13,Matthews13,Crespi15}, and to serve as a compact and flexible source for communication and computation protocols based on antisymmetric high-dimensional quantum states \cite{Jex03,Goyal14}.

\section{Theoretical framework}

\begin{figure}[h]
	\centering
	\includegraphics[width=\columnwidth]{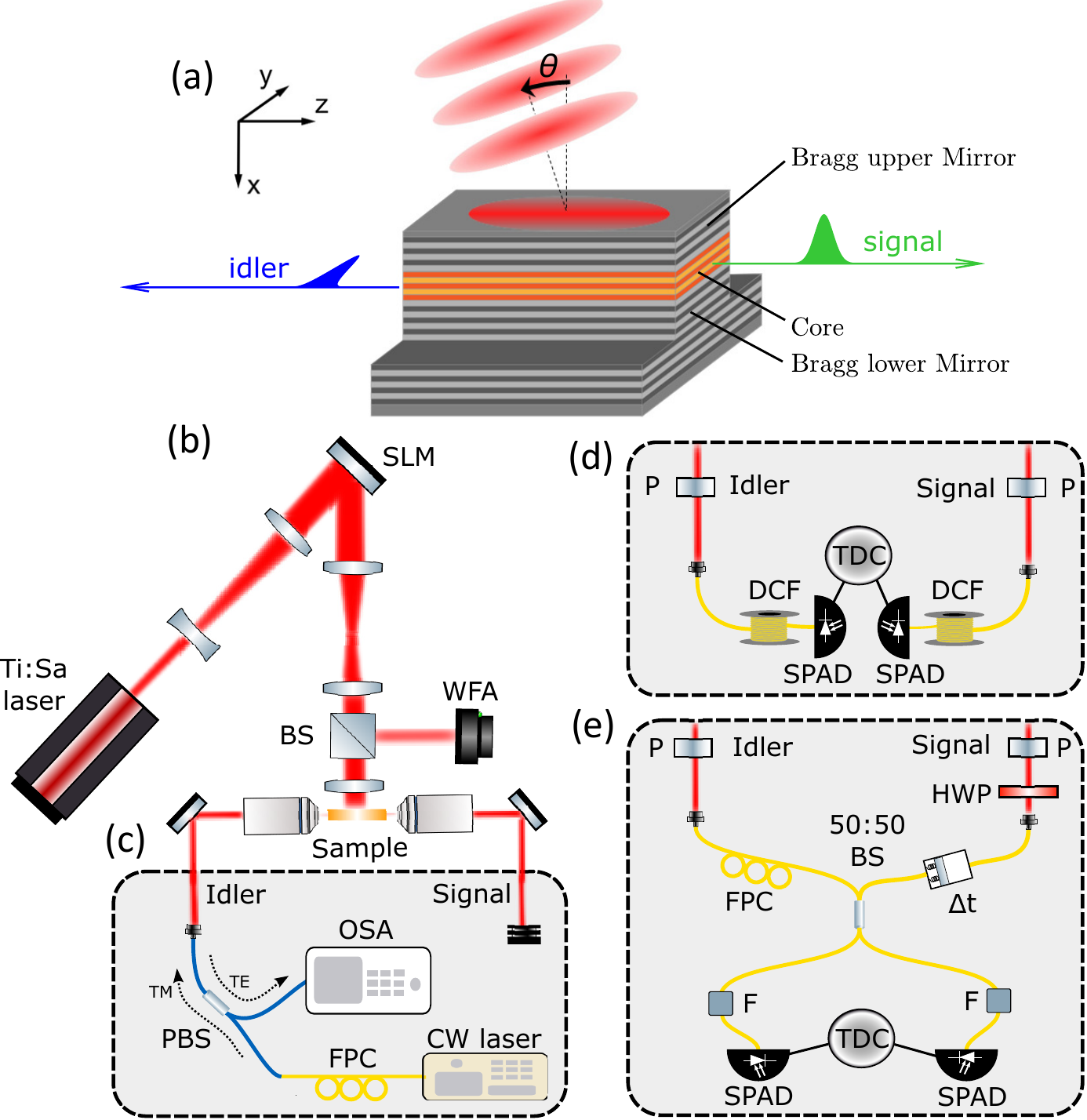}
	\caption{
		(a) Sketch of the AlGaAs ridge microcavity emitting photon pairs by PDC in a transverse pump geometry.
		(b)-(e) Sketch of the experiment, showing the pump shaping stage (b), stimulated emission tomography (c), fiber spectrograph (d) and Hong-Ou-Mandel (e) setups. Abbreviations: SLM=spatial light modulator, WFA=wavefront analyzer, PBS=polarizing beam splitter, FPC=fibered polarization controller, P=polarizer, HWP=half-wave plate, F=filter, DCF=dispersion compensating fiber, OSA=optical spectrum analyzer, SPAD=single-photon avalanche photodiode, TDC=time-to-digital converter.}
	\label{Fig1}
\end{figure}

The working principle of our semiconductor integrated source is sketched in Fig. \ref{Fig1}a. It is a Bragg ridge microcavity made of a stack of AlGaAs layers with alternating aluminum contents \cite{Caillet09,Orieux11,Orieux13}. The device is based on a transverse pump geometry, in which a pulsed pump laser beam impinging on top of the ridge (with an incidence angle $\theta$) generates pairs of counterpropagating, orthogonally polarized telecom-band photons (signal and idler) through PDC \cite{DeRossi02,Orieux13}. 
The Bragg mirrors provide both a vertical microcavity to enhance the pump field and a cladding for the twin-photon modes.
Of the two possible nonlinear interactions occurring in the device, in the following we consider the one that generates a TM-polarized signal photon (propagating along $z>0$, see Fig. \ref{Fig1}a) and a TE-polarized idler photon (propagating along $z<0$). The corresponding biphoton state reads
$\ket{\psi}= \iint d \omega_s d \omega_i {\rm JSA} (\omega_s,\omega_i)\hat{a}^\dagger_{s}(\omega_s) \hat{a}^\dagger_{i} (\omega_i)\ket{0,0}_{s,i}$,
where the operator $\hat{a}^\dagger_{s(i)}(\omega)$ creates a signal (idler) photon of frequency $\omega$.
The joint spectral amplitude JSA gives the probability amplitude of measuring a signal photon at frequency $\omega_s$ and an idler photon at frequency $\omega_i$. Neglecting group velocity dispersion (which is justified by the narrow spectral range of the generated photon pairs), and in the limit of narrow pump bandwidth, the JSA can be expressed as \cite{Boucher15,Barbieri17}:
\begin{equation}\label{Eq1}
{\rm JSA} (\omega_s,\omega_i)=\phi_{\rm spectral}(\omega_s+\omega_i) \, \phi_{\rm PM}(\omega_s-\omega_i)
\end{equation}
Here $\phi_{\rm spectral}$, reflecting the condition of energy conservation, corresponds to the spectrum of the pump beam and $\phi_{\rm PM}$, reflecting the phase-matching condition, is governed by the spatial properties of the pump beam:
\begin{equation}\label{Eq2}
\phi_{\rm PM}(\omega_s-\omega_i)=\int_{-L/2}^{L/2} dz \,\mathcal{A}_p (z) e^{-i (k_{\rm deg}+(\omega_s-\omega_i)/v_{\rm g})z}
\end{equation}
where $\mathcal{A}_p (z)$ is the pump amplitude profile along the waveguide direction, $L$ is the waveguide length, $v_g$ is the harmonic mean of the group velocities of the twin photon modes and $k_{\rm deg}=\omega_p\text{\rm sin}(\theta_{\rm deg})/c$. In the latter expression, $\omega_p$ is the pump central frequency, $c$ the light velocity and $\theta_{\rm deg}$ is the pump incidence angle needed to produce frequency-degenerate twin photons. Due to the small modal birefringence of our device ($\Delta n/n \sim 10^{-3}$), this degeneracy angle is slightly different from zero ($\theta_{\rm deg} \sim 0.5\degree$). When departing from this angle, the JSA gets translated in frequency space but its shape remains identical up to an excellent approximation \cite{Boucher15}.

\begin{figure}[h]
	\centering

\begingroup
\inputencoding{cp1252}%
\makeatletter
\providecommand\color[2][]{%
	\GenericError{(gnuplot) \space\space\space\@spaces}{%
		Package color not loaded in conjunction with
		terminal option `colourtext'%
	}{See the gnuplot documentation for explanation.%
	}{Either use 'blacktext' in gnuplot or load the package
		color.sty in LaTeX.}%
	\renewcommand\color[2][]{}%
}%
\providecommand\includegraphics[2][]{%
	\GenericError{(gnuplot) \space\space\space\@spaces}{%
		Package graphicx or graphics not loaded%
	}{See the gnuplot documentation for explanation.%
	}{The gnuplot epslatex terminal needs graphicx.sty or graphics.sty.}%
	\renewcommand\includegraphics[2][]{}%
}%
\providecommand\rotatebox[2]{#2}%
\@ifundefined{ifGPcolor}{%
	\newif\ifGPcolor
	\GPcolortrue
}{}%
\@ifundefined{ifGPblacktext}{%
	\newif\ifGPblacktext
	\GPblacktexttrue
}{}%
\let\gplgaddtomacro\g@addto@macro
\gdef\gplbacktext{}%
\gdef\gplfronttext{}%
\makeatother
\ifGPblacktext
\def\colorrgb#1{}%
\def\colorgray#1{}%
\else
\ifGPcolor
\def\colorrgb#1{\color[rgb]{#1}}%
\def\colorgray#1{\color[gray]{#1}}%
\expandafter\def\csname LTw\endcsname{\color{white}}%
\expandafter\def\csname LTb\endcsname{\color{black}}%
\expandafter\def\csname LTa\endcsname{\color{black}}%
\expandafter\def\csname LT0\endcsname{\color[rgb]{1,0,0}}%
\expandafter\def\csname LT1\endcsname{\color[rgb]{0,1,0}}%
\expandafter\def\csname LT2\endcsname{\color[rgb]{0,0,1}}%
\expandafter\def\csname LT3\endcsname{\color[rgb]{1,0,1}}%
\expandafter\def\csname LT4\endcsname{\color[rgb]{0,1,1}}%
\expandafter\def\csname LT5\endcsname{\color[rgb]{1,1,0}}%
\expandafter\def\csname LT6\endcsname{\color[rgb]{0,0,0}}%
\expandafter\def\csname LT7\endcsname{\color[rgb]{1,0.3,0}}%
\expandafter\def\csname LT8\endcsname{\color[rgb]{0.5,0.5,0.5}}%
\else
\def\colorrgb#1{\color{black}}%
\def\colorgray#1{\color[gray]{#1}}%
\expandafter\def\csname LTw\endcsname{\color{white}}%
\expandafter\def\csname LTb\endcsname{\color{black}}%
\expandafter\def\csname LTa\endcsname{\color{black}}%
\expandafter\def\csname LT0\endcsname{\color{black}}%
\expandafter\def\csname LT1\endcsname{\color{black}}%
\expandafter\def\csname LT2\endcsname{\color{black}}%
\expandafter\def\csname LT3\endcsname{\color{black}}%
\expandafter\def\csname LT4\endcsname{\color{black}}%
\expandafter\def\csname LT5\endcsname{\color{black}}%
\expandafter\def\csname LT6\endcsname{\color{black}}%
\expandafter\def\csname LT7\endcsname{\color{black}}%
\expandafter\def\csname LT8\endcsname{\color{black}}%
\fi
\fi
\setlength{\unitlength}{0.0500bp}%
\ifx\gptboxheight\undefined%
\newlength{\gptboxheight}%
\newlength{\gptboxwidth}%
\newsavebox{\gptboxtext}%
\fi%
\setlength{\fboxrule}{0.5pt}%
\setlength{\fboxsep}{1pt}%
\begin{picture}(4800.00,3280.00)%

\gplgaddtomacro\gplfronttext{%
	\csname LTb\endcsname
	\put(24,2402){\rotatebox{-270}{\makebox(0,0){\strut{}\normalsize{$\lambda_i$ (nm)}}}}%
	\csname LTb\endcsname
	\put(716,1655){\makebox(0,0){\strut{}\normalsize{$\lambda_s$ (nm})}}%
	\colorrgb{1.00,1.00,1.00}
	\put(161,1969){\rotatebox{-270}{\makebox(0,0){\strut{}\footnotesize $1549$}}}%
	\colorrgb{1.00,1.00,1.00}
	\put(161,2636){\rotatebox{-270}{\makebox(0,0){\strut{}\footnotesize $1550$}}}%
	\colorrgb{1.00,1.00,1.00}
	\put(216,1794){\makebox(0,0){\strut{}\footnotesize $1543$}}%
	\colorrgb{1.00,1.00,1.00}
	\put(883,1794){\makebox(0,0){\strut{}\footnotesize $1544$}}%
	\csname LTb\endcsname
	\put(1087,2733){\makebox(0,0){\strut{}\textcolor{white}{\normalsize{\textbf{(a)}}}}}%
	\csname LTb\endcsname
	\put(144,3246){\makebox(0,0)[l]{\strut{}\textcolor{red}{Experiment}}}%
	\csname LTb\endcsname
	\put(144,1443){\makebox(0,0)[l]{\strut{}\textcolor{red}{Theory}}}%
	\csname LTb\endcsname
	\put(717,3033){\makebox(0,0){\strut{}\normalsize{w=0.25 mm}}}%
	\csname LTb\endcsname
	\put(316,2042){\makebox(0,0)[l]{\strut{}\textcolor{white}{\footnotesize{K=1.34}}}}%
}%

\gplgaddtomacro\gplfronttext{%
	\csname LTb\endcsname
	\put(1862,1655){\makebox(0,0){\strut{}\normalsize{$\lambda_s$ (nm})}}%
	\colorrgb{1.00,1.00,1.00}
	\put(1306,2036){\rotatebox{-270}{\makebox(0,0){\strut{}\footnotesize $1549$}}}%
	\colorrgb{1.00,1.00,1.00}
	\put(1306,2704){\rotatebox{-270}{\makebox(0,0){\strut{}\footnotesize $1550$}}}%
	\colorrgb{1.00,1.00,1.00}
	\put(1428,1794){\makebox(0,0){\strut{}\footnotesize $1543$}}%
	\colorrgb{1.00,1.00,1.00}
	\put(2096,1794){\makebox(0,0){\strut{}\footnotesize $1544$}}%
	\csname LTb\endcsname
	\put(2233,2734){\makebox(0,0){\strut{}\textcolor{white}{\normalsize{\textbf{(b)}}}}}%
	\csname LTb\endcsname
	\put(1863,3033){\makebox(0,0){\strut{}\normalsize{w=0.4 mm}}}%
	\csname LTb\endcsname
	\put(1461,2042){\makebox(0,0)[l]{\strut{}\textcolor{white}{\footnotesize{K=1.18}}}}%
}%

\gplgaddtomacro\gplfronttext{%
	\csname LTb\endcsname
	\put(3008,1655){\makebox(0,0){\strut{}\normalsize{$\lambda_s$ (nm})}}%
	\colorrgb{1.00,1.00,1.00}
	\put(2453,2269){\rotatebox{-270}{\makebox(0,0){\strut{}\footnotesize $1550$}}}%
	\colorrgb{1.00,1.00,1.00}
	\put(2775,1794){\makebox(0,0){\strut{}\footnotesize $1544$}}%
	\colorrgb{1.00,1.00,1.00}
	\put(3442,1794){\makebox(0,0){\strut{}\footnotesize $1545$}}%
	\csname LTb\endcsname
	\put(3379,2733){\makebox(0,0){\strut{}\textcolor{white}{\normalsize{\textbf{(c)}}}}}%
	\csname LTb\endcsname
	\put(3008,3033){\makebox(0,0){\strut{}\normalsize{w=0.6 mm}}}%
	\csname LTb\endcsname
	\put(2608,2042){\makebox(0,0)[l]{\strut{}\textcolor{white}{\footnotesize{K=1.06}}}}%
}%

\gplgaddtomacro\gplfronttext{%
	\csname LTb\endcsname
	\put(4154,1655){\makebox(0,0){\strut{}\normalsize{$\lambda_s$ (nm})}}%
	\colorrgb{1.00,1.00,1.00}
	\put(3599,2269){\rotatebox{-270}{\makebox(0,0){\strut{}\footnotesize $1550$}}}%
	\colorrgb{1.00,1.00,1.00}
	\put(3888,1794){\makebox(0,0){\strut{}\footnotesize $1544$}}%
	\colorrgb{1.00,1.00,1.00}
	\put(4555,1794){\makebox(0,0){\strut{}\footnotesize $1545$}}%
	\colorrgb{1.00,1.00,1.00}
	\put(4781,1902){\makebox(0,0)[l]{\strut{}\footnotesize $0$}}%
	\colorrgb{1.00,1.00,1.00}
	\put(4781,2903){\makebox(0,0)[l]{\strut{}\footnotesize $1$}}%
	\csname LTb\endcsname
	\put(4525,2733){\makebox(0,0){\strut{}\textcolor{white}{\normalsize{\textbf{(d)}}}}}%
	\csname LTb\endcsname
	\put(4154,3033){\makebox(0,0){\strut{}\normalsize{w=1 mm}}}%
	\csname LTb\endcsname
	\put(3754,2042){\makebox(0,0)[l]{\strut{}\textcolor{white}{\footnotesize{K=1.12}}}}%
}%

\gplgaddtomacro\gplfronttext{%
	\csname LTb\endcsname
	\put(24,828){\rotatebox{-270}{\makebox(0,0){\strut{}\normalsize{$\lambda_i$ (nm)}}}}%
	\csname LTb\endcsname
	\put(716,81){\makebox(0,0){\strut{}\normalsize{$\lambda_s$ (nm})}}%
	\colorrgb{1.00,1.00,1.00}
	\put(161,395){\rotatebox{-270}{\makebox(0,0){\strut{}\footnotesize $1549$}}}%
	\colorrgb{1.00,1.00,1.00}
	\put(161,1062){\rotatebox{-270}{\makebox(0,0){\strut{}\footnotesize $1550$}}}%
	\colorrgb{1.00,1.00,1.00}
	\put(216,220){\makebox(0,0){\strut{}\footnotesize $1543$}}%
	\colorrgb{1.00,1.00,1.00}
	\put(883,220){\makebox(0,0){\strut{}\footnotesize $1544$}}%
	\csname LTb\endcsname
	\put(1087,1159){\makebox(0,0){\strut{}\textcolor{white}{\normalsize{\textbf{(e)}}}}}%
	\csname LTb\endcsname
	\put(316,468){\makebox(0,0)[l]{\strut{}\textcolor{white}{\footnotesize{K=1.31}}}}%
}%

\gplgaddtomacro\gplfronttext{%
	\csname LTb\endcsname
	\put(1862,81){\makebox(0,0){\strut{}\normalsize{$\lambda_s$ (nm})}}%
	\colorrgb{1.00,1.00,1.00}
	\put(1306,462){\rotatebox{-270}{\makebox(0,0){\strut{}\footnotesize $1549$}}}%
	\colorrgb{1.00,1.00,1.00}
	\put(1306,1130){\rotatebox{-270}{\makebox(0,0){\strut{}\footnotesize $1550$}}}%
	\colorrgb{1.00,1.00,1.00}
	\put(1428,220){\makebox(0,0){\strut{}\footnotesize $1543$}}%
	\colorrgb{1.00,1.00,1.00}
	\put(2096,220){\makebox(0,0){\strut{}\footnotesize $1544$}}%
	\csname LTb\endcsname
	\put(2233,1160){\makebox(0,0){\strut{}\textcolor{white}{\normalsize{\textbf{(f)}}}}}%
	\csname LTb\endcsname
	\put(1461,468){\makebox(0,0)[l]{\strut{}\textcolor{white}{\footnotesize{K=1.04}}}}%
}%

\gplgaddtomacro\gplfronttext{%
	\csname LTb\endcsname
	\put(3008,81){\makebox(0,0){\strut{}\normalsize{$\lambda_s$ (nm})}}%
	\colorrgb{1.00,1.00,1.00}
	\put(2453,695){\rotatebox{-270}{\makebox(0,0){\strut{}\footnotesize $1550$}}}%
	\colorrgb{1.00,1.00,1.00}
	\put(2775,220){\makebox(0,0){\strut{}\footnotesize $1544$}}%
	\colorrgb{1.00,1.00,1.00}
	\put(3442,220){\makebox(0,0){\strut{}\footnotesize $1545$}}%
	\csname LTb\endcsname
	\put(3379,1159){\makebox(0,0){\strut{}\textcolor{white}{\normalsize{\textbf{(g)}}}}}%
	\csname LTb\endcsname
	\put(2608,468){\makebox(0,0)[l]{\strut{}\textcolor{white}{\footnotesize{K=1.01}}}}%
}%

\gplgaddtomacro\gplfronttext{%
	\csname LTb\endcsname
	\put(4154,81){\makebox(0,0){\strut{}\normalsize{$\lambda_s$ (nm})}}%
	\colorrgb{1.00,1.00,1.00}
	\put(3599,695){\rotatebox{-270}{\makebox(0,0){\strut{}\footnotesize $1550$}}}%
	\colorrgb{1.00,1.00,1.00}
	\put(3888,220){\makebox(0,0){\strut{}\footnotesize $1544$}}%
	\colorrgb{1.00,1.00,1.00}
	\put(4555,220){\makebox(0,0){\strut{}\footnotesize $1545$}}%
	\colorrgb{1.00,1.00,1.00}
	\put(4781,328){\makebox(0,0)[l]{\strut{}\footnotesize $0$}}%
	\colorrgb{1.00,1.00,1.00}
	\put(4781,1329){\makebox(0,0)[l]{\strut{}\footnotesize $1$}}%
	\csname LTb\endcsname
	\put(4525,1159){\makebox(0,0){\strut{}\textcolor{white}{\normalsize{\textbf{(h)}}}}}%
	\csname LTb\endcsname
	\put(3754,468){\makebox(0,0)[l]{\strut{}\textcolor{white}{\footnotesize{K=1.12}}}}%
}%
\gplbacktext
\put(0,0){\includegraphics{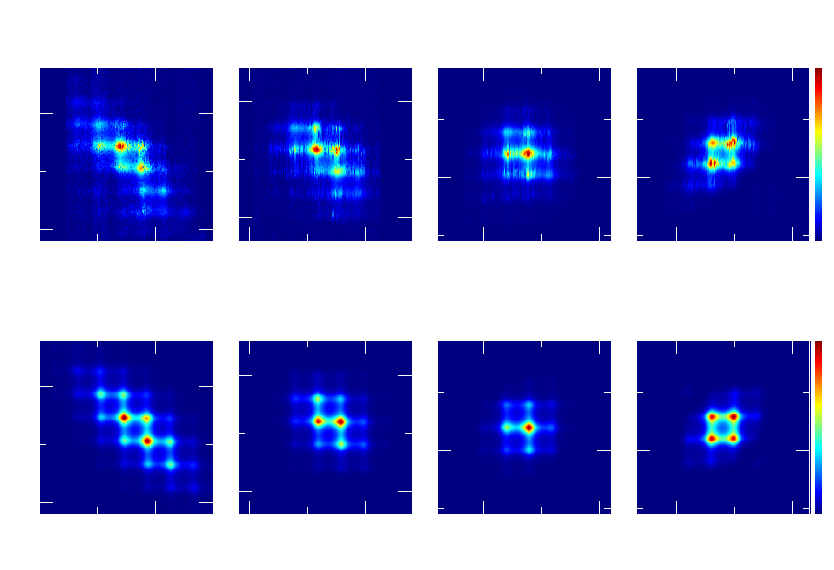}}%
\gplfronttext
\end{picture}%
\endgroup

\caption{
	Measured joint spectral intensities (JSI) for increasing values of the pump beam waist: (a) $0.25$ mm, (b) $0.4$ mm, (c) $0.6$ mm and (d) $1$ mm. (e)-(h) Numerically simulated JSI for the above parameters. $\lambda_s$ and $\lambda_i$ denote the wavelength of the signal and idler photons, respectively.
}
	\label{Fig2}
\end{figure}

\begin{figure*}[t]
	\centering
\begingroup
\inputencoding{cp1252}%
\makeatletter
\providecommand\color[2][]{%
	\GenericError{(gnuplot) \space\space\space\@spaces}{%
		Package color not loaded in conjunction with
		terminal option `colourtext'%
	}{See the gnuplot documentation for explanation.%
	}{Either use 'blacktext' in gnuplot or load the package
		color.sty in LaTeX.}%
	\renewcommand\color[2][]{}%
}%
\providecommand\includegraphics[2][]{%
	\GenericError{(gnuplot) \space\space\space\@spaces}{%
		Package graphicx or graphics not loaded%
	}{See the gnuplot documentation for explanation.%
	}{The gnuplot epslatex terminal needs graphicx.sty or graphics.sty.}%
	\renewcommand\includegraphics[2][]{}%
}%
\providecommand\rotatebox[2]{#2}%
\@ifundefined{ifGPcolor}{%
	\newif\ifGPcolor
	\GPcolortrue
}{}%
\@ifundefined{ifGPblacktext}{%
	\newif\ifGPblacktext
	\GPblacktexttrue
}{}%
\let\gplgaddtomacro\g@addto@macro
\gdef\gplbacktext{}%
\gdef\gplfronttext{}%
\makeatother
\ifGPblacktext
\def\colorrgb#1{}%
\def\colorgray#1{}%
\else
\ifGPcolor
\def\colorrgb#1{\color[rgb]{#1}}%
\def\colorgray#1{\color[gray]{#1}}%
\expandafter\def\csname LTw\endcsname{\color{white}}%
\expandafter\def\csname LTb\endcsname{\color{black}}%
\expandafter\def\csname LTa\endcsname{\color{black}}%
\expandafter\def\csname LT0\endcsname{\color[rgb]{1,0,0}}%
\expandafter\def\csname LT1\endcsname{\color[rgb]{0,1,0}}%
\expandafter\def\csname LT2\endcsname{\color[rgb]{0,0,1}}%
\expandafter\def\csname LT3\endcsname{\color[rgb]{1,0,1}}%
\expandafter\def\csname LT4\endcsname{\color[rgb]{0,1,1}}%
\expandafter\def\csname LT5\endcsname{\color[rgb]{1,1,0}}%
\expandafter\def\csname LT6\endcsname{\color[rgb]{0,0,0}}%
\expandafter\def\csname LT7\endcsname{\color[rgb]{1,0.3,0}}%
\expandafter\def\csname LT8\endcsname{\color[rgb]{0.5,0.5,0.5}}%
\else
\def\colorrgb#1{\color{black}}%
\def\colorgray#1{\color[gray]{#1}}%
\expandafter\def\csname LTw\endcsname{\color{white}}%
\expandafter\def\csname LTb\endcsname{\color{black}}%
\expandafter\def\csname LTa\endcsname{\color{black}}%
\expandafter\def\csname LT0\endcsname{\color{black}}%
\expandafter\def\csname LT1\endcsname{\color{black}}%
\expandafter\def\csname LT2\endcsname{\color{black}}%
\expandafter\def\csname LT3\endcsname{\color{black}}%
\expandafter\def\csname LT4\endcsname{\color{black}}%
\expandafter\def\csname LT5\endcsname{\color{black}}%
\expandafter\def\csname LT6\endcsname{\color{black}}%
\expandafter\def\csname LT7\endcsname{\color{black}}%
\expandafter\def\csname LT8\endcsname{\color{black}}%
\fi
\fi
\setlength{\unitlength}{0.0500bp}%
\ifx\gptboxheight\undefined%
\newlength{\gptboxheight}%
\newlength{\gptboxwidth}%
\newsavebox{\gptboxtext}%
\fi%
\setlength{\fboxrule}{0.5pt}%
\setlength{\fboxsep}{1pt}%
\begin{picture}(10480.00,3680.00)%
\gplgaddtomacro\gplbacktext{%
}%
\gplgaddtomacro\gplfronttext{%
	\csname LTb\endcsname
	\put(105,3311){\makebox(0,0)[l]{\strut{}\textcolor{black}{\normalsize{\textbf{(a)}}}}}%
	\csname LTb\endcsname
	\put(2201,3642){\makebox(0,0)[l]{\strut{}\textcolor{red}{Experiment}}}%
	\csname LTb\endcsname
	\put(2201,1674){\makebox(0,0)[l]{\strut{}\textcolor{red}{Theory}}}%
	\csname LTb\endcsname
	\put(419,2575){\makebox(0,0){\normalsize{Pump}}}%
	\csname LTb\endcsname
	\put(597,1398){\makebox(0,0)[l]{\strut{}\textcolor{white}{\normalsize{Waveguide}}}}%
	\csname LTb\endcsname
	\put(629,3017){\makebox(0,0)[l]{\strut{}\normalsize{Phase Shift $\Delta\varphi$}}}%
	\csname LTb\endcsname
	\put(996,2759){\makebox(0,0)[l]{\strut{}\normalsize{$\theta$}}}%
}%

\gplgaddtomacro\gplfronttext{%
	\csname LTb\endcsname
	\put(2244,2759){\rotatebox{-270}{\makebox(0,0){\strut{}\normalsize{$\lambda_i$ (nm)}}}}%
	\csname LTb\endcsname
	\put(3082,1857){\makebox(0,0){\strut{}\normalsize{$\lambda_s$ (nm})}}%
	\colorrgb{1.00,1.00,1.00}
	\put(2387,2301){\rotatebox{-270}{\makebox(0,0){\strut{}\footnotesize $1549.5$}}}%
	\colorrgb{1.00,1.00,1.00}
	\put(2387,3134){\rotatebox{-270}{\makebox(0,0){\strut{}\footnotesize $1550.5$}}}%
	\colorrgb{1.00,1.00,1.00}
	\put(2754,2011){\makebox(0,0){\strut{}\footnotesize $1543.5$}}%
	\colorrgb{1.00,1.00,1.00}
	\put(3587,2011){\makebox(0,0){\strut{}\footnotesize $1544.5$}}%
	\csname LTb\endcsname
	\put(3550,3197){\makebox(0,0){\strut{}\textcolor{white}{\normalsize{\textbf{(b)}}}}}%
	\csname LTb\endcsname
	\put(3128,3477){\makebox(0,0){\strut{}\normalsize{$\Delta\varphi=0$}}}%
}%

\gplgaddtomacro\gplfronttext{%
	\csname LTb\endcsname
	\put(3852,2759){\rotatebox{-270}{\makebox(0,0){\strut{}\normalsize{$\lambda_i$ (nm)}}}}%
	\csname LTb\endcsname
	\put(4690,1857){\makebox(0,0){\strut{}\normalsize{$\lambda_s$ (nm})}}%
	\colorrgb{1.00,1.00,1.00}
	\put(3995,2259){\rotatebox{-270}{\makebox(0,0){\strut{}\footnotesize $1549.5$}}}%
	\colorrgb{1.00,1.00,1.00}
	\put(3995,3092){\rotatebox{-270}{\makebox(0,0){\strut{}\footnotesize $1550.5$}}}%
	\colorrgb{1.00,1.00,1.00}
	\put(4320,2011){\makebox(0,0){\strut{}\footnotesize $1543.5$}}%
	\colorrgb{1.00,1.00,1.00}
	\put(5153,2011){\makebox(0,0){\strut{}\footnotesize $1544.5$}}%
	\csname LTb\endcsname
	\put(5158,3197){\makebox(0,0){\strut{}\textcolor{white}{\normalsize{\textbf{(c)}}}}}%
	\csname LTb\endcsname
	\put(4663,3477){\makebox(0,0){\strut{}\normalsize{$\Delta\varphi= \pi/4$}}}%
}%

\gplgaddtomacro\gplfronttext{%
	\csname LTb\endcsname
	\put(5460,2759){\rotatebox{-270}{\makebox(0,0){\strut{}\normalsize{$\lambda_i$ (nm)}}}}%
	\csname LTb\endcsname
	\put(6298,1857){\makebox(0,0){\strut{}\normalsize{$\lambda_s$ (nm})}}%
	\colorrgb{1.00,1.00,1.00}
	\put(5603,2176){\rotatebox{-270}{\makebox(0,0){\strut{}\footnotesize $1549.5$}}}%
	\colorrgb{1.00,1.00,1.00}
	\put(5603,3009){\rotatebox{-270}{\makebox(0,0){\strut{}\footnotesize $1550.5$}}}%
	\colorrgb{1.00,1.00,1.00}
	\put(5928,2011){\makebox(0,0){\strut{}\footnotesize $1543.5$}}%
	\colorrgb{1.00,1.00,1.00}
	\put(6761,2011){\makebox(0,0){\strut{}\footnotesize $1544.5$}}%
	\csname LTb\endcsname
	\put(6766,3197){\makebox(0,0){\strut{}\textcolor{white}{\normalsize{\textbf{(d)}}}}}%
	\csname LTb\endcsname
	\put(6287,3477){\makebox(0,0){\strut{}\normalsize{$\Delta\varphi=\pi/2$}}}%
}%

\gplgaddtomacro\gplfronttext{%
	\csname LTb\endcsname
	\put(7067,2759){\rotatebox{-270}{\makebox(0,0){\strut{}\normalsize{$\lambda_i$ (nm)}}}}%
	\csname LTb\endcsname
	\put(7905,1857){\makebox(0,0){\strut{}\normalsize{$\lambda_s$ (nm})}}%
	\colorrgb{1.00,1.00,1.00}
	\put(7210,2926){\rotatebox{-270}{\makebox(0,0){\strut{}\footnotesize $1550.5$}}}%
	\colorrgb{1.00,1.00,1.00}
	\put(7210,3342){\rotatebox{-270}{\makebox(0,0){\strut{}\footnotesize $1551$}}}%
	\colorrgb{1.00,1.00,1.00}
	\put(7535,2011){\makebox(0,0){\strut{}\footnotesize $1543.5$}}%
	\colorrgb{1.00,1.00,1.00}
	\put(8368,2011){\makebox(0,0){\strut{}\footnotesize $1544.5$}}%
	\csname LTb\endcsname
	\put(8373,3197){\makebox(0,0){\strut{}\textcolor{white}{\normalsize{\textbf{(e)}}}}}%
	\csname LTb\endcsname
	\put(7912,3477){\makebox(0,0){\strut{}\normalsize{$\Delta\varphi=3\pi/2$}}}%
}%

\gplgaddtomacro\gplfronttext{%
	\csname LTb\endcsname
	\put(8675,2759){\rotatebox{-270}{\makebox(0,0){\strut{}\normalsize{$\lambda_i$ (nm)}}}}%
	\csname LTb\endcsname
	\put(9513,1857){\makebox(0,0){\strut{}\normalsize{$\lambda_s$ (nm})}}%
	\colorrgb{1.00,1.00,1.00}
	\put(8818,2884){\rotatebox{-270}{\makebox(0,0){\strut{}\footnotesize $1550.5$}}}%
	\colorrgb{1.00,1.00,1.00}
	\put(8818,3301){\rotatebox{-270}{\makebox(0,0){\strut{}\footnotesize $1551$}}}%
	\colorrgb{1.00,1.00,1.00}
	\put(9143,2011){\makebox(0,0){\strut{}\footnotesize $1543.5$}}%
	\colorrgb{1.00,1.00,1.00}
	\put(9976,2011){\makebox(0,0){\strut{}\footnotesize $1544.5$}}%
	\colorrgb{1.00,1.00,1.00}
	\put(10287,2134){\makebox(0,0)[l]{\strut{}\footnotesize $0$}}%
	\colorrgb{1.00,1.00,1.00}
	\put(10287,3384){\makebox(0,0)[l]{\strut{}\footnotesize $1$}}%
	\csname LTb\endcsname
	\put(9981,3197){\makebox(0,0){\strut{}\textcolor{white}{\normalsize{\textbf{(f)}}}}}%
	\csname LTb\endcsname
	\put(9578,3477){\makebox(0,0){\strut{}\normalsize{$\Delta\varphi=\pi$}}}%
}%

\gplgaddtomacro\gplfronttext{%
	\csname LTb\endcsname
	\put(2244,930){\rotatebox{-270}{\makebox(0,0){\strut{}\normalsize{$\lambda_i$ (nm)}}}}%
	\csname LTb\endcsname
	\put(3093,17){\makebox(0,0){\strut{}\normalsize{$\lambda_s$ (nm})}}%
	\colorrgb{1.00,1.00,1.00}
	\put(2387,464){\rotatebox{-270}{\makebox(0,0){\strut{}\footnotesize $1549.5$}}}%
	\colorrgb{1.00,1.00,1.00}
	\put(2387,1312){\rotatebox{-270}{\makebox(0,0){\strut{}\footnotesize $1550.5$}}}%
	\colorrgb{1.00,1.00,1.00}
	\put(2759,171){\makebox(0,0){\strut{}\footnotesize $1543.5$}}%
	\colorrgb{1.00,1.00,1.00}
	\put(3607,171){\makebox(0,0){\strut{}\footnotesize $1544.5$}}%
	\csname LTb\endcsname
	\put(3569,1375){\makebox(0,0){\strut{}\textcolor{white}{\normalsize{\textbf{(g)}}}}}%
	\csname LTb\endcsname
	\put(2564,421){\makebox(0,0)[l]{\strut{}\textcolor{white}{\footnotesize{K=1.37}}}}%
}%

\gplgaddtomacro\gplfronttext{%
	\csname LTb\endcsname
	\put(3852,930){\rotatebox{-270}{\makebox(0,0){\strut{}\normalsize{$\lambda_i$ (nm)}}}}%
	\csname LTb\endcsname
	\put(4701,17){\makebox(0,0){\strut{}\normalsize{$\lambda_s$ (nm})}}%
	\colorrgb{1.00,1.00,1.00}
	\put(3995,421){\rotatebox{-270}{\makebox(0,0){\strut{}\footnotesize $1549.5$}}}%
	\colorrgb{1.00,1.00,1.00}
	\put(3995,1269){\rotatebox{-270}{\makebox(0,0){\strut{}\footnotesize $1550.5$}}}%
	\colorrgb{1.00,1.00,1.00}
	\put(4324,171){\makebox(0,0){\strut{}\footnotesize $1543.5$}}%
	\colorrgb{1.00,1.00,1.00}
	\put(5172,171){\makebox(0,0){\strut{}\footnotesize $1544.5$}}%
	\csname LTb\endcsname
	\put(5177,1375){\makebox(0,0){\strut{}\textcolor{white}{\normalsize{\textbf{(h)}}}}}%
	\csname LTb\endcsname
	\put(4172,421){\makebox(0,0)[l]{\strut{}\textcolor{white}{\footnotesize{K=1.44}}}}%
}%

\gplgaddtomacro\gplfronttext{%
	\csname LTb\endcsname
	\put(5460,929){\rotatebox{-270}{\makebox(0,0){\strut{}\normalsize{$\lambda_i$ (nm)}}}}%
	\csname LTb\endcsname
	\put(6308,17){\makebox(0,0){\strut{}\normalsize{$\lambda_s$ (nm})}}%
	\colorrgb{1.00,1.00,1.00}
	\put(5603,336){\rotatebox{-270}{\makebox(0,0){\strut{}\footnotesize $1549.5$}}}%
	\colorrgb{1.00,1.00,1.00}
	\put(5603,1184){\rotatebox{-270}{\makebox(0,0){\strut{}\footnotesize $1550.5$}}}%
	\colorrgb{1.00,1.00,1.00}
	\put(5932,171){\makebox(0,0){\strut{}\footnotesize $1543.5$}}%
	\colorrgb{1.00,1.00,1.00}
	\put(6780,171){\makebox(0,0){\strut{}\footnotesize $1544.5$}}%
	\csname LTb\endcsname
	\put(6784,1374){\makebox(0,0){\strut{}\textcolor{white}{\normalsize{\textbf{(i)}}}}}%
	\csname LTb\endcsname
	\put(5780,421){\makebox(0,0)[l]{\strut{}\textcolor{white}{\footnotesize{K=1.65}}}}%
}%

\gplgaddtomacro\gplfronttext{%
	\csname LTb\endcsname
	\put(7067,930){\rotatebox{-270}{\makebox(0,0){\strut{}\normalsize{$\lambda_i$ (nm)}}}}%
	\csname LTb\endcsname
	\put(7916,17){\makebox(0,0){\strut{}\normalsize{$\lambda_s$ (nm})}}%
	\colorrgb{1.00,1.00,1.00}
	\put(7210,1100){\rotatebox{-270}{\makebox(0,0){\strut{}\footnotesize $1550.5$}}}%
	\colorrgb{1.00,1.00,1.00}
	\put(7210,1524){\rotatebox{-270}{\makebox(0,0){\strut{}\footnotesize $1551$}}}%
	\colorrgb{1.00,1.00,1.00}
	\put(7539,171){\makebox(0,0){\strut{}\footnotesize $1543.5$}}%
	\colorrgb{1.00,1.00,1.00}
	\put(8387,171){\makebox(0,0){\strut{}\footnotesize $1544.5$}}%
	\csname LTb\endcsname
	\put(8392,1375){\makebox(0,0){\strut{}\textcolor{white}{\normalsize{\textbf{(j)}}}}}%
	\csname LTb\endcsname
	\put(7387,421){\makebox(0,0)[l]{\strut{}\textcolor{white}{\footnotesize{K=2.00}}}}%
}%

\gplgaddtomacro\gplfronttext{%
	\csname LTb\endcsname
	\put(8675,929){\rotatebox{-270}{\makebox(0,0){\strut{}\normalsize{$\lambda_i$ (nm)}}}}%
	\csname LTb\endcsname
	\put(9523,17){\makebox(0,0){\strut{}\normalsize{$\lambda_s$ (nm})}}%
	\colorrgb{1.00,1.00,1.00}
	\put(8818,1057){\rotatebox{-270}{\makebox(0,0){\strut{}\footnotesize $1550.5$}}}%
	\colorrgb{1.00,1.00,1.00}
	\put(8818,1480){\rotatebox{-270}{\makebox(0,0){\strut{}\footnotesize $1551$}}}%
	\colorrgb{1.00,1.00,1.00}
	\put(9147,171){\makebox(0,0){\strut{}\footnotesize $1543.5$}}%
	\colorrgb{1.00,1.00,1.00}
	\put(9995,171){\makebox(0,0){\strut{}\footnotesize $1544.5$}}%
	\colorrgb{1.00,1.00,1.00}
	\put(10310,294){\makebox(0,0)[l]{\strut{}\footnotesize $0$}}%
	\colorrgb{1.00,1.00,1.00}
	\put(10310,1565){\makebox(0,0)[l]{\strut{}\footnotesize $1$}}%
	\csname LTb\endcsname
	\put(9999,1374){\makebox(0,0){\strut{}\textcolor{white}{\normalsize{\textbf{(k)}}}}}%
	\csname LTb\endcsname
	\put(8995,421){\makebox(0,0)[l]{\strut{}\textcolor{white}{\footnotesize{K=2.24}}}}%
}%
\gplbacktext
\put(0,0){\includegraphics{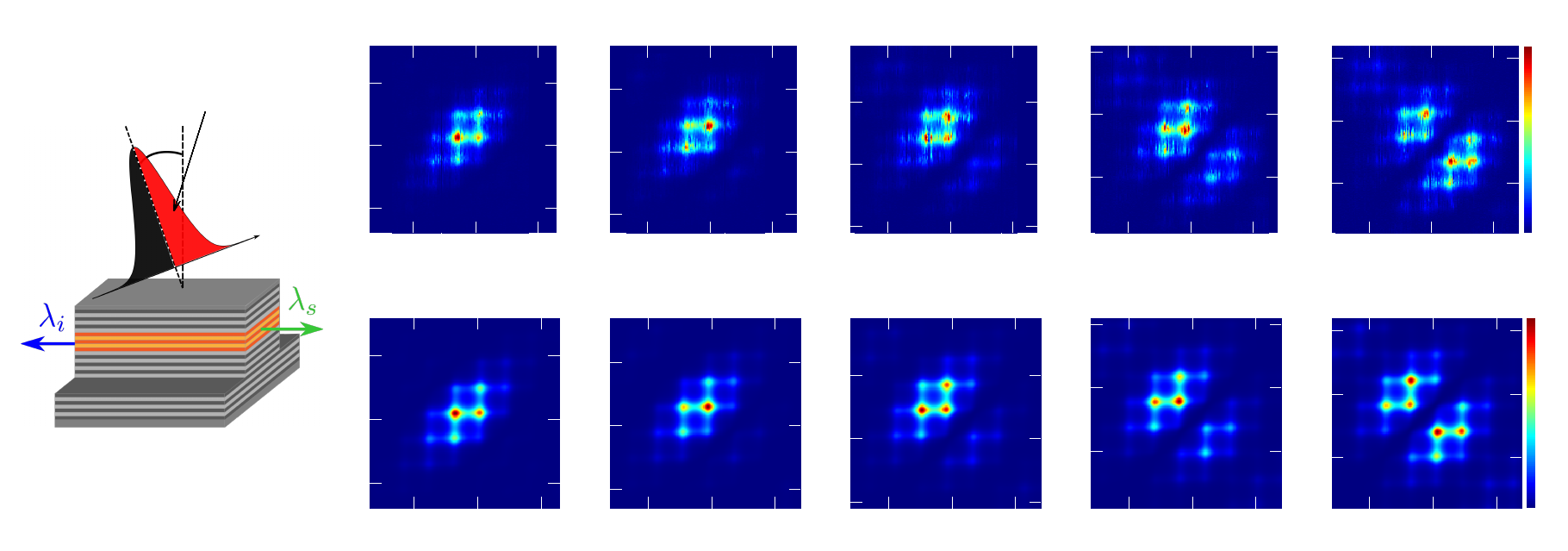}}%
\gplfronttext
\end{picture}%
\endgroup

\caption{
	(a) Sketch of the pumping geometry to control the symmetry of the biphoton quantum frequency state.
	(b)-(f) Measured JSI for increasing values of the phase step $\Delta \varphi$ between the two halves of the pump beam.
	(g)-(k) Corresponding simulated JSI.
}

\label{Fig3}
\end{figure*}

Equation (1) indicates that the shape of the JSA along the diagonal direction of the biphoton frequency space ($\omega_+=\omega_s+\omega_i$) and that along the antidiagonal direction ($\omega_-=\omega_s-\omega_i$) can be tuned independently, by varying respectively the spectral or spatial properties of the pump beam, providing a simple and versatile means to engineer the frequency-time correlations of the photon pairs \cite{Boucher15}. In addition, in contrast to the co-propopagative regime of guided-wave PDC \cite{Kumar13,Ansari18} the signal and idler photons are here produced in two distinct spatial modes, facilitating their further utilization in protocols. Here, we will exploit the spatial control of the pump beam in intensity and phase by using a spatial light modulator (SLM).

\section{Experimental setup}

The experimental setup is shown in Fig. \ref{Fig1}b. The AlGaAs source (ridge length $L=2$ mm, width 6 $\mu$m, height 7 $\mu$m) is pumped with a pulsed Ti:Sa laser with wavelength $\lambda_p \simeq 773 $ nm, pulse duration $\simeq$ 6 ps, repetition rate 76 MHz and average pump power $50$ mW incident on the sample. The pump beam is shaped in intensity and phase using a reflective phase-only SLM (Holoeye Leto) in a 4f configuration, and analyzed with a Wavefront Analyser (WFA) to verify the obtained modulation. Finally a cylindrical lens focuses the beam on the top of the waveguide, and the PDC photons are collected with two microscope objectives and collimated into telecom optical fibers. To characterize the emitted quantum states we measure the Joint Spectral Intensity (JSI), which is the modulus squared of the JSA, by using a Stimulated Emission Tomography (SET) technique \cite{Eckstein14} as sketched in Fig. \ref{Fig1}c. In this technique, in addition to the transverse pump beam, a TM polarized CW telecom laser (seed beam), injected through one facet of the waveguide, stimulates the generation of the (TE polarized) idler field by difference frequency generation, and its spectrum is recorded with an Optical Spectrum Analyzer (OSA). The wavelength of the seed laser is swept so as to iteratively reconstruct the whole JSI.

\section{Control of frequency correlations}

We first demonstrate the control over frequency correlations by varying the spatial extension of the pump beam. 
We pump the device with Gaussian pump profiles, $\mathcal{A}_p (z)=e^{-z^2/w^2} e^{ik z}$, where $w$ is the beam waist on the waveguide and $k=\omega_p\sin(\theta)/c$ is the projection of the pump wavevector along the $z$ direction.
In this situation, the phase-matching term $\phi_{\rm PM}(\omega_s-\omega_i)$ is real and corresponds, in the biphoton frequency space $(\omega_s,\omega_i)$, to a stripe aligned along the diagonal, with a width inversely proportional to the pump waist (in the limit where $L \gg w$).
The other term of the JSA, $\phi_{\rm spectral}(\omega_s+\omega_i)$, is given by the spectral distribution of the pump beam: since we use unchirped (Fourier-transform limited) pulses, it is also a real function and corresponds to a stripe aligned along the antidiagonal, with a width inversely proportional to the duration of the pump pulses.
The JSA is the product of these two functions: it thus has the shape of an ellipse whose size and orientation is determined by the pump waist and pulse duration.

Fig. \ref{Fig2}a reports the JSI measured by the SET technique for a pump waist $w=0.25$ mm and a pulse duration of $6$ ps; the pump angle $\theta$ is slightly offset from degeneracy as required for the SET measurement \cite{Eckstein14}.
The spectrum is aligned along the antidiagonal, corresponding to a frequency-anticorrelated state. 
We note the presence of a grid-like pattern, which is related to the reflectivity of the waveguide facets: this creates a  Fabry-Perot cavity along the $z$ direction, whose transmission resonances modulate the joint spectrum \cite{Eckstein14}. 
This effect could be exploited to facilitate the manipulation of the frequency degree of freedom by discretizing it, as is the case for quantum frequency combs \cite{Roslund14,Kues19,Maltese20}; on the other hand, it could be removed if needed by depositing an anti-reflection coating e.g. in silicon nitride \cite{Apiratikul14}.
Starting from the anticorrelated spectrum of Fig. \ref{Fig2}a, \ref{Fig2}b-d shows the JSI measured for increasing values of the pump waist. 
We observe that the extension of the JSI along the antidiagonal direction progressively shrinks, transforming the initial state into a frequency-correlated state when $w=1$ mm (Fig. \ref{Fig2}d). For the intermediate value $w=0.6$ mm (Fig. \ref{Fig2}c), the width of the phase-matching and spectral terms of the JSA are nearly equal, yielding a circular joint spectrum corresponding to a frequency-separable state.
The numerical simulations in Fig. \ref{Fig2}e-h, which take into account modal birefringence, chromatic dispersion and cavity effects in the sample are in excellent agreement with the experiment. We show also on each panel the calculated Schmidt number $K$ (obtained from the JSA), which quantifies the effective number of orthogonal frequency modes spanned by the biphoton wavefunction \cite{Mosley08}. For the experimental data (Fig. \ref{Fig2}a-d) the Schmidt number is determined by assuming a flat-phase JSA, a reasonable approximation here since we use unchirped pulses with flat spatial phase profiles.
The Schmidt number initially decreases, reaches $K\simeq 1$ (corresponding to a separable state) when the JSI is circular, before increasing again when the state becomes frequency-correlated. Note that quantum states with higher Schmidt numbers (i.e. involving more time-frequency modes) could be obtained with the same source by tuning the pumping parameters (see Supplementary Information for a quantitative discussion).

Overall, the results presented in Fig. \ref{Fig2} demonstrate a flexible frequency engineering of biphoton quantum states, which can be exploited to adapt the AlGaAs integrated source to different quantum information applications requiring either anticorrelated \cite{Kues17}, separable \cite{Mosley08} or correlated frequency states \cite{Giovannetti01,Lutz14}. In contrast to filtering approaches that decrease the source brightness by removing unwanted parts of the spectrum \cite{Lu07,Kues19}, here the full biphoton spectral intensity is entirely directed into the desired shape at the generation stage. The pair production rate is here $\simeq 10$ MHz at the chip output, corresponding to a brightness of $\simeq 200$ kHz/mW.

\section{Control of wavefunction symmetry and exchange statistics}

We now investigate further control of the quantum frequency state by engineering the phase profile of the pump beam.
A first natural way is to impose a phase step $\Delta \varphi$ between the two halves of the pump spot, as sketched in Fig. \ref{Fig3}a. Placing the pump spot at the center $z=0$ of the waveguide, the pump amplitude profile reads $\mathcal{A}_p (z)=F(z) e^{-z^2/w^2}e^{ik z}$, with $F(z)=1$ for $z<0$ and $F(z)=e^{i \Delta \varphi}$ for $z>0$. 
When pumping at the degeneracy angle $\theta_{\rm deg}$, one can show that the phase-matching term (Eq. 2) takes the form (see Supplementary Information):
\begin{equation}
\phi_{\rm PM}(\omega_s,\omega_i)=f(\omega_s,\omega_i)+e^{i \Delta \varphi} f(\omega_i,\omega_s)
\end{equation}\label{Eq3}%
with $f(\omega_s,\omega_i)=\int_0^{L/2} dz \,e^{-z^2/w^2} e^{i (\omega_s-\omega_i)z/v_{\rm g}}$. 
As can be directly read from Eq. (3), for $\Delta \varphi=0$ (which corresponds to a standard Gaussian beam as studied previously) the phase-matching function is symmetric with respect to particle exchange ($\phi_{\rm PM}(\omega_s,\omega_i)=\phi_{\rm PM}(\omega_i,\omega_s)$), while for $\Delta \varphi= \pi$ it becomes antisymmetric ($\phi_{\rm PM}(\omega_s,\omega_i)=-\phi_{\rm PM}(\omega_i,\omega_s)$). Since the spectral function $\phi_{\rm spectral}$ is always symmetric (it depends only on the frequency sum $\omega_s+\omega_i$), the parity of $\phi_{\rm PM}$ directly translates to the JSA. This analysis thus predicts that a simple phase engineering of the pump beam should allow to control the symmetry of the spectral wavefunction of the photon pairs.

We experimentally implement this concept and show in Fig. \ref{Fig3}b-f the measured JSI for increasing values of the phase step $\Delta \varphi$, at fixed pump waist ($1$ mm, as in Fig. \ref{Fig2}d) and pulse duration ($4$ ps). 
Starting from a frequency-correlated state at  $\Delta \varphi=0$ we observe the progressive appearance of a second lobe in the joint spectrum as $\Delta \varphi$ increases. These results are in good agreement with the numerical simulations (Fig. \ref{Fig3}g-k), where we show also the calculated Schmidt numbers (here the non-flat phase structure of the JSA does not allow to  determine $K$ experimentally).
When $\Delta \varphi=\pi$ (Fig. \ref{Fig3}f,k) the spectrum is split into two lobes of equal intensity, and vanishes along the diagonal axis between the two lobes. According to the previous theoretical analysis, there is a $\pi$ offset between the spectral phase of points which are mirror-symmetric with respect to this diagonal axis. However the JSI measurement is not sensitive to such phase information: to retrieve this information and probe the biphoton spectral wavefunction parity we will exploit two-photon interference in a Hong-Ou-Mandel (HOM) experiment.

The experimental HOM setup is shown in Fig. \ref{Fig1}e. The polarization of the signal photon is rotated and aligned with that of the idler, then the signal photon enters a fibered delay line, before recombining with the idler on a fibered 50/50 beamsplitter. Coincidence counts at the outputs (after a long-wave pass filter to remove luminescence noise) are monitored while scanning the delay $\tau$ of the interferometer. This HOM experiment has in principle 4 possible outcomes: the two photons can either leave the beamsplitter through the same output port (bunching) or through different ports (antibunching) -- with 2 possibilities in each case. When the entangled state is symmetric, antibunching probability amplitudes cancel each other, leaving only bunching events; when the biphoton state is antisymmetric, the reverse scenario occurs, leaving only antibunching events as would be the case for (independent) fermions \cite{Walborn03,Fedrizzi09,Sansoni10,Zhang16}. 

\begin{figure*}[t]
	\centering
	\begingroup
	\inputencoding{cp1252}%
	\makeatletter
	\providecommand\color[2][]{%
		\GenericError{(gnuplot) \space\space\space\@spaces}{%
			Package color not loaded in conjunction with
			terminal option `colourtext'%
		}{See the gnuplot documentation for explanation.%
		}{Either use 'blacktext' in gnuplot or load the package
			color.sty in LaTeX.}%
		\renewcommand\color[2][]{}%
	}%
	\providecommand\includegraphics[2][]{%
		\GenericError{(gnuplot) \space\space\space\@spaces}{%
			Package graphicx or graphics not loaded%
		}{See the gnuplot documentation for explanation.%
		}{The gnuplot epslatex terminal needs graphicx.sty or graphics.sty.}%
		\renewcommand\includegraphics[2][]{}%
	}%
	\providecommand\rotatebox[2]{#2}%
	\@ifundefined{ifGPcolor}{%
		\newif\ifGPcolor
		\GPcolortrue
	}{}%
	\@ifundefined{ifGPblacktext}{%
		\newif\ifGPblacktext
		\GPblacktexttrue
	}{}%
	\let\gplgaddtomacro\g@addto@macro
	\gdef\gplbacktext{}%
	\gdef\gplfronttext{}%
	\makeatother
	\ifGPblacktext
	\def\colorrgb#1{}%
	\def\colorgray#1{}%
	\else
	\ifGPcolor
	\def\colorrgb#1{\color[rgb]{#1}}%
	\def\colorgray#1{\color[gray]{#1}}%
	\expandafter\def\csname LTw\endcsname{\color{white}}%
	\expandafter\def\csname LTb\endcsname{\color{black}}%
	\expandafter\def\csname LTa\endcsname{\color{black}}%
	\expandafter\def\csname LT0\endcsname{\color[rgb]{1,0,0}}%
	\expandafter\def\csname LT1\endcsname{\color[rgb]{0,1,0}}%
	\expandafter\def\csname LT2\endcsname{\color[rgb]{0,0,1}}%
	\expandafter\def\csname LT3\endcsname{\color[rgb]{1,0,1}}%
	\expandafter\def\csname LT4\endcsname{\color[rgb]{0,1,1}}%
	\expandafter\def\csname LT5\endcsname{\color[rgb]{1,1,0}}%
	\expandafter\def\csname LT6\endcsname{\color[rgb]{0,0,0}}%
	\expandafter\def\csname LT7\endcsname{\color[rgb]{1,0.3,0}}%
	\expandafter\def\csname LT8\endcsname{\color[rgb]{0.5,0.5,0.5}}%
	\else
	\def\colorrgb#1{\color{black}}%
	\def\colorgray#1{\color[gray]{#1}}%
	\expandafter\def\csname LTw\endcsname{\color{white}}%
	\expandafter\def\csname LTb\endcsname{\color{black}}%
	\expandafter\def\csname LTa\endcsname{\color{black}}%
	\expandafter\def\csname LT0\endcsname{\color{black}}%
	\expandafter\def\csname LT1\endcsname{\color{black}}%
	\expandafter\def\csname LT2\endcsname{\color{black}}%
	\expandafter\def\csname LT3\endcsname{\color{black}}%
	\expandafter\def\csname LT4\endcsname{\color{black}}%
	\expandafter\def\csname LT5\endcsname{\color{black}}%
	\expandafter\def\csname LT6\endcsname{\color{black}}%
	\expandafter\def\csname LT7\endcsname{\color{black}}%
	\expandafter\def\csname LT8\endcsname{\color{black}}%
	\fi
	\fi
	\setlength{\unitlength}{0.0500bp}%
	\ifx\gptboxheight\undefined%
	\newlength{\gptboxheight}%
	\newlength{\gptboxwidth}%
	\newsavebox{\gptboxtext}%
	\fi%
	\setlength{\fboxrule}{0.5pt}%
	\setlength{\fboxsep}{1pt}%
	\begin{picture}(9620.00,3400.00)%
	
	\gplgaddtomacro\gplfronttext{%
		\csname LTb\endcsname
		\put(128,2688){\rotatebox{-270}{\makebox(0,0){\strut{}\normalsize{$\lambda_i$ (nm)}}}}%
		\csname LTb\endcsname
		\put(1226,1769){\makebox(0,0){\strut{}\normalsize{$\lambda_s$ (nm})}}%
		\colorrgb{1.00,1.00,1.00}
		\put(563,2076){\makebox(0,0)[r]{\strut{}\footnotesize $1545$}}%
		\colorrgb{1.00,1.00,1.00}
		\put(563,2576){\makebox(0,0)[r]{\strut{}\footnotesize $1546$}}%
		\colorrgb{1.00,1.00,1.00}
		\put(563,3075){\makebox(0,0)[r]{\strut{}\footnotesize $1547$}}%
		\colorrgb{1.00,1.00,1.00}
		\put(614,1938){\makebox(0,0){\strut{}\footnotesize $1545$}}%
		\colorrgb{1.00,1.00,1.00}
		\put(1114,1938){\makebox(0,0){\strut{}\footnotesize $1546$}}%
		\colorrgb{1.00,1.00,1.00}
		\put(1614,1938){\makebox(0,0){\strut{}\footnotesize $1547$}}%
		\colorrgb{1.00,1.00,1.00}
		\put(1982,2076){\makebox(0,0)[l]{\strut{}\footnotesize $0$}}%
		\colorrgb{1.00,1.00,1.00}
		\put(1982,3300){\makebox(0,0)[l]{\strut{}\footnotesize $1$}}%
		\csname LTb\endcsname
		\put(1680,3080){\makebox(0,0){\strut{}\textcolor{white}{\normalsize{\textbf{(a)}}}}}%
		\csname LTb\endcsname
		\put(896,3116){\makebox(0,0){\strut{}\textcolor{white}{\normalsize{Exp.}}}}%
	}%
	
	\gplgaddtomacro\gplfronttext{%
		\csname LTb\endcsname
		\put(2576,2707){\rotatebox{-270}{\makebox(0,0){\strut{}\normalsize{Coincidences (/s)}}}}%
		\csname LTb\endcsname
		\put(3790,1769){\makebox(0,0){\strut{}\normalsize{delay $\tau$ (ps)}}}%
		\colorrgb{0.00,0.00,0.00}
		\put(2910,2115){\makebox(0,0)[r]{\strut{}\footnotesize $0$}}%
		\colorrgb{0.00,0.00,0.00}
		\put(2910,2411){\makebox(0,0)[r]{\strut{}\footnotesize $25$}}%
		\colorrgb{0.00,0.00,0.00}
		\put(2910,2708){\makebox(0,0)[r]{\strut{}\footnotesize $50$}}%
		\colorrgb{0.00,0.00,0.00}
		\put(2910,3004){\makebox(0,0)[r]{\strut{}\footnotesize $75$}}%
		\colorrgb{0.00,0.00,0.00}
		\put(2910,3300){\makebox(0,0)[r]{\strut{}\footnotesize $100$}}%
		\colorrgb{0.00,0.00,0.00}
		\put(3226,1938){\makebox(0,0){\strut{}\footnotesize $-20$}}%
		\colorrgb{0.00,0.00,0.00}
		\put(3508,1938){\makebox(0,0){\strut{}\footnotesize $-10$}}%
		\colorrgb{0.00,0.00,0.00}
		\put(3791,1938){\makebox(0,0){\strut{}\footnotesize $0$}}%
		\colorrgb{0.00,0.00,0.00}
		\put(4073,1938){\makebox(0,0){\strut{}\footnotesize $10$}}%
		\colorrgb{0.00,0.00,0.00}
		\put(4355,1938){\makebox(0,0){\strut{}\footnotesize $20$}}%
		\csname LTb\endcsname
		\put(4423,3122){\makebox(0,0){\strut{}\textcolor{black}{\normalsize{\textbf{(b)}}}}}%
		\csname LTb\endcsname
		\put(3269,3122){\makebox(0,0){\strut{}\textcolor{black}{\normalsize{Exp.}}}}%
	}%
	
	\gplgaddtomacro\gplfronttext{%
		\csname LTb\endcsname
		\put(4941,2707){\rotatebox{-270}{\makebox(0,0){\strut{}\normalsize{\shortstack{Coincidence \\ probability}}}}}%
		\csname LTb\endcsname
		\put(6195,1769){\makebox(0,0){\strut{}\normalsize{delay $\tau$ (ps)}}}%
		\colorrgb{0.00,0.00,0.00}
		\put(5315,2115){\makebox(0,0)[r]{\strut{}\footnotesize $0$}}%
		\colorrgb{0.00,0.00,0.00}
		\put(5315,2708){\makebox(0,0)[r]{\strut{}\footnotesize $0.5$}}%
		\colorrgb{0.00,0.00,0.00}
		\put(5315,3300){\makebox(0,0)[r]{\strut{}\footnotesize $1$}}%
		\colorrgb{0.00,0.00,0.00}
		\put(5631,1938){\makebox(0,0){\strut{}\footnotesize $-20$}}%
		\colorrgb{0.00,0.00,0.00}
		\put(5913,1938){\makebox(0,0){\strut{}\footnotesize $-10$}}%
		\colorrgb{0.00,0.00,0.00}
		\put(6196,1938){\makebox(0,0){\strut{}\footnotesize $0$}}%
		\colorrgb{0.00,0.00,0.00}
		\put(6478,1938){\makebox(0,0){\strut{}\footnotesize $10$}}%
		\colorrgb{0.00,0.00,0.00}
		\put(6760,1938){\makebox(0,0){\strut{}\footnotesize $20$}}%
		\csname LTb\endcsname
		\put(6828,3122){\makebox(0,0){\strut{}\textcolor{black}{\normalsize{\textbf{(c)}}}}}%
		\csname LTb\endcsname
		\put(5674,3122){\makebox(0,0){\strut{}\textcolor{black}{Th.}}}%
	}%
	
	\gplgaddtomacro\gplfronttext{%
		\csname LTb\endcsname
		\put(7410,2707){\rotatebox{-270}{\makebox(0,0){\strut{}\normalsize{$\lambda_s - \lambda_i$ (nm)}}}}%
		\csname LTb\endcsname
		\put(8493,1769){\makebox(0,0){\strut{}\normalsize{$t_-$ (ps)}}}%
		\colorrgb{0.00,0.00,0.00}
		\put(7662,2337){\makebox(0,0)[r]{\strut{}\footnotesize $-1$}}%
		\colorrgb{0.00,0.00,0.00}
		\put(7662,2708){\makebox(0,0)[r]{\strut{}\footnotesize $0$}}%
		\colorrgb{0.00,0.00,0.00}
		\put(7662,3078){\makebox(0,0)[r]{\strut{}\footnotesize $1$}}%
		\colorrgb{0.00,0.00,0.00}
		\put(7964,1938){\makebox(0,0){\strut{}\footnotesize $-20$}}%
		\colorrgb{0.00,0.00,0.00}
		\put(8228,1938){\makebox(0,0){\strut{}\footnotesize $-10$}}%
		\colorrgb{0.00,0.00,0.00}
		\put(8493,1938){\makebox(0,0){\strut{}\footnotesize $0$}}%
		\colorrgb{0.00,0.00,0.00}
		\put(8758,1938){\makebox(0,0){\strut{}\footnotesize $10$}}%
		\colorrgb{0.00,0.00,0.00}
		\put(9022,1938){\makebox(0,0){\strut{}\footnotesize $20$}}%
		\colorrgb{0.00,0.00,0.00}
		\put(9396,2115){\makebox(0,0)[l]{\strut{}\footnotesize $-1$}}%
		\colorrgb{0.00,0.00,0.00}
		\put(9396,2707){\makebox(0,0)[l]{\strut{}\footnotesize $0$}}%
		\colorrgb{0.00,0.00,0.00}
		\put(9396,3300){\makebox(0,0)[l]{\strut{}\footnotesize $1$}}%
		\csname LTb\endcsname
		\put(9086,3122){\makebox(0,0){\strut{}\textcolor{black}{\normalsize{\textbf{(d)}}}}}%
		\csname LTb\endcsname
		\put(8004,3122){\makebox(0,0){\strut{}\textcolor{black}{\normalsize{Th.}}}}%
	}%
	
	\gplgaddtomacro\gplfronttext{%
		\csname LTb\endcsname
		\put(128,988){\rotatebox{-270}{\makebox(0,0){\strut{}\normalsize{$\lambda_i$ (nm)}}}}%
		\csname LTb\endcsname
		\put(1226,69){\makebox(0,0){\strut{}\normalsize{$\lambda_s$ (nm})}}%
		\colorrgb{1.00,1.00,1.00}
		\put(563,376){\makebox(0,0)[r]{\strut{}\footnotesize $1545$}}%
		\colorrgb{1.00,1.00,1.00}
		\put(563,876){\makebox(0,0)[r]{\strut{}\footnotesize $1546$}}%
		\colorrgb{1.00,1.00,1.00}
		\put(563,1376){\makebox(0,0)[r]{\strut{}\footnotesize $1547$}}%
		\colorrgb{1.00,1.00,1.00}
		\put(614,238){\makebox(0,0){\strut{}\footnotesize $1545$}}%
		\colorrgb{1.00,1.00,1.00}
		\put(1114,238){\makebox(0,0){\strut{}\footnotesize $1546$}}%
		\colorrgb{1.00,1.00,1.00}
		\put(1614,238){\makebox(0,0){\strut{}\footnotesize $1547$}}%
		\colorrgb{1.00,1.00,1.00}
		\put(1982,376){\makebox(0,0)[l]{\strut{}\footnotesize $0$}}%
		\colorrgb{1.00,1.00,1.00}
		\put(1982,1601){\makebox(0,0)[l]{\strut{}\footnotesize $1$}}%
		\csname LTb\endcsname
		\put(1680,1381){\makebox(0,0){\strut{}\textcolor{white}{\normalsize{\textbf{(e)}}}}}%
		\csname LTb\endcsname
		\put(896,1417){\makebox(0,0){\strut{}\textcolor{white}{\normalsize{Exp.}}}}%
	}%
	
	\gplgaddtomacro\gplfronttext{%
		\csname LTb\endcsname
		\put(2601,1008){\rotatebox{-270}{\makebox(0,0){\strut{}\normalsize{Coincidences (/s)}}}}%
		\csname LTb\endcsname
		\put(3765,69){\makebox(0,0){\strut{}\normalsize{delay $\tau$ (ps)}}}%
		\colorrgb{0.00,0.00,0.00}
		\put(2884,415){\makebox(0,0)[r]{\strut{}\footnotesize $0$}}%
		\colorrgb{0.00,0.00,0.00}
		\put(2884,786){\makebox(0,0)[r]{\strut{}\footnotesize $25$}}%
		\colorrgb{0.00,0.00,0.00}
		\put(2884,1156){\makebox(0,0)[r]{\strut{}\footnotesize $50$}}%
		\colorrgb{0.00,0.00,0.00}
		\put(2884,1527){\makebox(0,0)[r]{\strut{}\footnotesize $75$}}%
		\colorrgb{0.00,0.00,0.00}
		\put(3200,238){\makebox(0,0){\strut{}\footnotesize $-20$}}%
		\colorrgb{0.00,0.00,0.00}
		\put(3483,238){\makebox(0,0){\strut{}\footnotesize $-10$}}%
		\colorrgb{0.00,0.00,0.00}
		\put(3765,238){\makebox(0,0){\strut{}\footnotesize $0$}}%
		\colorrgb{0.00,0.00,0.00}
		\put(4048,238){\makebox(0,0){\strut{}\footnotesize $10$}}%
		\colorrgb{0.00,0.00,0.00}
		\put(4330,238){\makebox(0,0){\strut{}\footnotesize $20$}}%
		\csname LTb\endcsname
		\put(4398,1423){\makebox(0,0){\strut{}\textcolor{black}{\normalsize{\textbf{(f)}}}}}%
		\csname LTb\endcsname
		\put(3243,1423){\makebox(0,0){\strut{}\textcolor{black}{\normalsize{Exp.}}}}%
	}%
	
	\gplgaddtomacro\gplfronttext{%
		\csname LTb\endcsname
		\put(4941,1008){\rotatebox{-270}{\makebox(0,0){\strut{}\normalsize{\shortstack{Coincidence \\ probability}}}}}%
		\csname LTb\endcsname
		\put(6195,69){\makebox(0,0){\strut{}\normalsize{delay $\tau$ (ps)}}}%
		\colorrgb{0.00,0.00,0.00}
		\put(5315,415){\makebox(0,0)[r]{\strut{}\footnotesize $0$}}%
		\colorrgb{0.00,0.00,0.00}
		\put(5315,954){\makebox(0,0)[r]{\strut{}\footnotesize $0.5$}}%
		\colorrgb{0.00,0.00,0.00}
		\put(5315,1493){\makebox(0,0)[r]{\strut{}\footnotesize $1$}}%
		\colorrgb{0.00,0.00,0.00}
		\put(5631,238){\makebox(0,0){\strut{}\footnotesize $-20$}}%
		\colorrgb{0.00,0.00,0.00}
		\put(5913,238){\makebox(0,0){\strut{}\footnotesize $-10$}}%
		\colorrgb{0.00,0.00,0.00}
		\put(6196,238){\makebox(0,0){\strut{}\footnotesize $0$}}%
		\colorrgb{0.00,0.00,0.00}
		\put(6478,238){\makebox(0,0){\strut{}\footnotesize $10$}}%
		\colorrgb{0.00,0.00,0.00}
		\put(6760,238){\makebox(0,0){\strut{}\footnotesize $20$}}%
		\csname LTb\endcsname
		\put(6828,1423){\makebox(0,0){\strut{}\textcolor{black}{\normalsize{\textbf{(g)}}}}}%
		\csname LTb\endcsname
		\put(5674,1423){\makebox(0,0){\strut{}\textcolor{black}{\normalsize{Th.}}}}%
	}%
	
	\gplgaddtomacro\gplfronttext{%
		\csname LTb\endcsname
		\put(7410,1008){\rotatebox{-270}{\makebox(0,0){\strut{}\normalsize{$\lambda_s - \lambda_i$ (nm)}}}}%
		\csname LTb\endcsname
		\put(8493,69){\makebox(0,0){\strut{}\normalsize{$t_-$ (ps)}}}%
		\colorrgb{0.00,0.00,0.00}
		\put(7662,637){\makebox(0,0)[r]{\strut{}\footnotesize $-1$}}%
		\colorrgb{0.00,0.00,0.00}
		\put(7662,1008){\makebox(0,0)[r]{\strut{}\footnotesize $0$}}%
		\colorrgb{0.00,0.00,0.00}
		\put(7662,1379){\makebox(0,0)[r]{\strut{}\footnotesize $1$}}%
		\colorrgb{0.00,0.00,0.00}
		\put(7964,238){\makebox(0,0){\strut{}\footnotesize $-20$}}%
		\colorrgb{0.00,0.00,0.00}
		\put(8228,238){\makebox(0,0){\strut{}\footnotesize $-10$}}%
		\colorrgb{0.00,0.00,0.00}
		\put(8493,238){\makebox(0,0){\strut{}\footnotesize $0$}}%
		\colorrgb{0.00,0.00,0.00}
		\put(8758,238){\makebox(0,0){\strut{}\footnotesize $10$}}%
		\colorrgb{0.00,0.00,0.00}
		\put(9022,238){\makebox(0,0){\strut{}\footnotesize $20$}}%
		\colorrgb{0.00,0.00,0.00}
		\put(9396,415){\makebox(0,0)[l]{\strut{}\footnotesize $-1$}}%
		\colorrgb{0.00,0.00,0.00}
		\put(9396,1008){\makebox(0,0)[l]{\strut{}\footnotesize $0$}}%
		\colorrgb{0.00,0.00,0.00}
		\put(9396,1601){\makebox(0,0)[l]{\strut{}\footnotesize $1$}}%
		\csname LTb\endcsname
		\put(9086,1423){\makebox(0,0){\strut{}\textcolor{black}{\normalsize{\textbf{(h)}}}}}%
		\csname LTb\endcsname
		\put(8004,1423){\makebox(0,0){\strut{}\textcolor{black}{\normalsize{Th.}}}}%
	}%
	\gplbacktext
	\put(0,0){\includegraphics{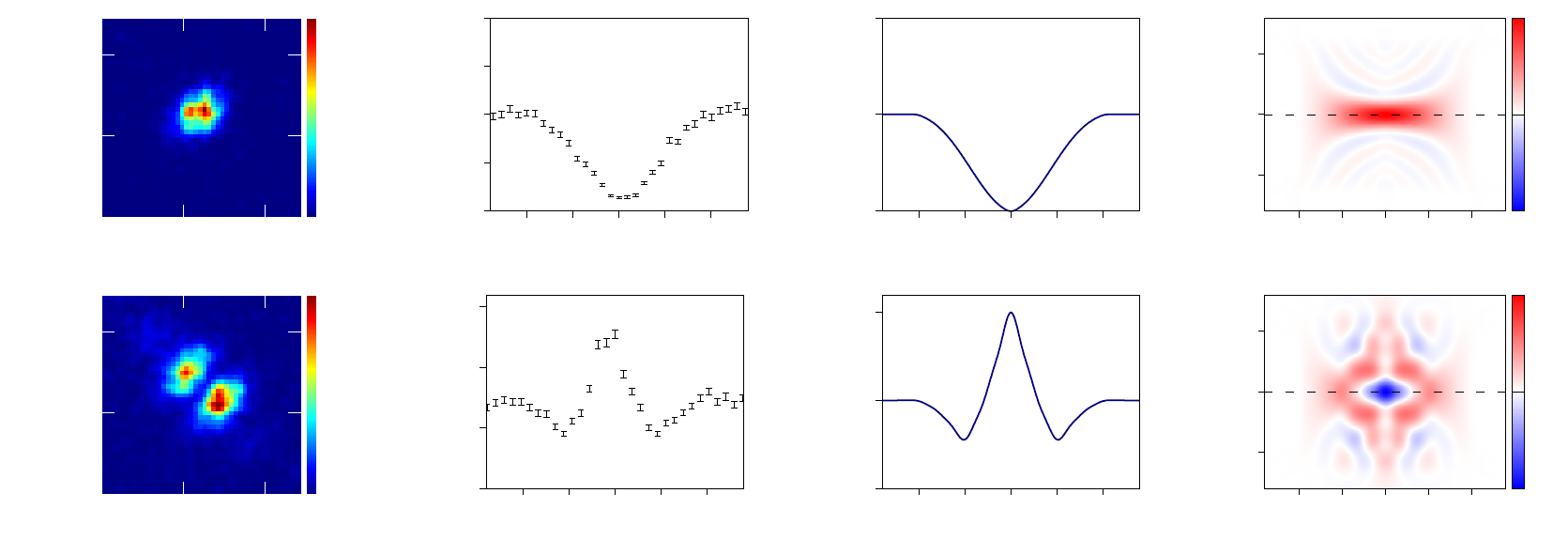}}%
	\gplfronttext
	\end{picture}%
	\endgroup
	
\caption{
	(a) Measured JSI for a Gaussian pump beam, leading to a symmetric frequency-entangled state. 
	(b) Corresponding measured and (c) calculated coincidences in a Hong-Ou-Mandel experiment, and (d) calculated chronocyclic Wigner function $W_{-}$ (normalized so that $\pm1$ corresponds to a HOM dip (peak) of full visibility).
	(e-h) Same as (a-d) but when applying a $\pi$ phase step at the center of the pump beam, leading to an antisymmetric frequency-entangled state. Experimental data correspond to raw (uncorrected)  coincidence counts.
}
	\label{Fig4}
\end{figure*}

We first consider the quantum frequency state obtained when pumping the waveguide with a Gaussian of flat phase profile ($\Delta \varphi=0$). Fig. \ref{Fig4}a shows the corresponding JSI measured at degeneracy with a fiber spectrograph \cite{Eckstein14} (see Fig. \ref{Fig1}d): each photon of the pairs is sent into a spool of highly dispersive fiber so as to convert the frequency information into a time-of-arrival information, which is recorded with single-photon avalanche photodiodes (SPAD, of detection efficiency $25\%$) connected to a time-to-digital converter (TDC). This technique has here a lower resolution ($\Delta \lambda \sim 200$ pm) than the SET technique but contrary to the latter, it can be implemented at frequency degeneracy.
The result of the HOM experiment performed with this quantum state is shown in Fig.~\ref{Fig4}b, with the corresponding simulation in Fig. \ref{Fig4}c. We observe a coincidence dip (i.e. two-photon bunching), confirming the symmetric nature of the frequency state. The experimental dip visibility, defined as $V=(N_{\infty}-N_{0})/N_{\infty}$, with $N_{\infty}$ ($N_{0}$) the mean coincidence counts at long (zero) time delay, is 88 $\%$ (using raw coincidence counts); our simulations indicate that this value is mainly limited by slight imperfections of the pump spatial profile and incidence angle (see Supplementary Information).

We next consider the biphoton state obtained when imposing a phase step $\Delta \varphi=\pi$ at the center of the pump spot, resulting in a split JSI as seen in the spectrum of Fig. \ref{Fig4}e, measured at frequency degeneracy. Here, the HOM interferogram (Fig. \ref{Fig4}f-g) shows a coincidence peak (antibunching), demonstrating the antisymmetric nature of the frequency state and the effectively fermionic behavior of the photons. The raw experimental visibility is here of 77~$\%$, again mainly limited by pump imperfections; the side dips at $\pm12$ ps delay are due to the specific shape of the joint spectrum.

Interestingly, the anti-bunching behavior evidenced for the antisymmetric frequency state (Fig. \ref{Fig4}f) is a direct proof of entanglement \cite{Eckstein08,Fedrizzi09}, and more precisely, of entanglement with non-Gaussian statistics \cite{Gomes09,Douce13} in the continuous variables formed by the time-frequency degrees of freedom of the biphotons. 
This non-Gaussian entanglement is associated to the negativity of the chronocyclic Wigner function (CWF) \cite{Brecht13}, $W(\omega_s,\omega_i,t_s,t_i)$, which gives the quasi-probability amplitude of measuring a signal photon at frequency $\omega_s$ and time $t_s$ and an idler photon at frequency $\omega_i$ and time $t_i$. Similarly to the JSA (\eqref{Eq1}), in our case the CWF can be factorized into a spectral and a phase-matching contributions, $W=W_{+}(\omega_+,t_+) \, W_{-}(\omega_-,t_-)$, with $\omega_{\pm}=\omega_s \pm \omega_i$ and $t_{\pm}=(t_s \pm t_i)/2$. The coincidence probability $P(\tau)$ in the HOM experiment is determined by the cut of the $W_{-}$ function along $\omega_-=0$ (see dotted lines in Fig. 4d,h), $P(\tau)=\frac{1}{2}\left(1-W_{-}(0,\tau)\right)$ \cite{Douce13,Tischler15}.
Fig. \ref{Fig4}d-h shows the $W_{-}$ function calculated for our symmetric and antisymmetric frequency states, respectively. In the latter case the CWF takes negative values (reaching the theoretical minimum of  $-1$) at $\omega_-=0$ (i.e. $\lambda_s-\lambda_i=0$), evidencing non-Gaussian entanglement. Note that in Fig. \ref{Fig4}d, a small negativity ($\sim -0.05$) also appears at non-zero values of $\omega_-$ due to the finite length of the device.
A full experimental determination of the CWF could be performed by using a generalized HOM experiment, where a frequency shift is added between the two photons (using eg. an electro-optic modulator) in addition to the usual temporal delay. Measuring the HOM trace for various frequency shifts would then allow to move along the vertical axis of the CWF shown in Fig. \ref{Fig4}d-h and reconstruct the $W_{-}$  function slice by slice \cite{Douce13,Tischler15}: this provides an alternative and promising route to the characterization of a quantum frequency state that does not require a direct measurement of the phase of the JSA. Interestingly, in the particular case of our counter-propagative source it has been shown that instead of using an electro-optic modulator, a simple change of the pump incidence angle can be used to scan the Wigner function along the frequency difference axis \cite{Boucher15}.

\section{Discussion}

In summary, we have demonstrated a flexible control over the spectral wavefunction and particle statistics of photon pairs, with a chip-integrated source and directly at the generation stage.
The symmetry control of high-dimensional entangled states has been demonstrated previously in the spatial degree of freedom \cite{Walborn03, Zhang16}, but using bulk sources only. 
In the frequency degree of freedom, displaying strong potential for applications thanks to its robustness to propagation and capability to convey large-scale quantum information into a single spatial mode, a recent work demonstrated the integrated and post-manipulation-free control of the spectrum of biphotons by engineering the spectrum of the pump field, leading in particular to the production of time-frequency Bell states and the implementation of high-dimensional operations in the time-frequency domain \cite{Ansari18}.
Another work developed a method to control two-color entanglement and gain control over the biphoton spectrum \cite{Jin18}, but this approach requires two passages in a bulk source and post-manipulation with a dispersive element, and is limited to the production of two-color entangled states. By contrast, here we experimentally demonstrate a general method providing a complete toolbox to engineer quantum frequency states, at the generation stage and using a chip-based source: these features are essential in view of practical and scalable applications for quantum information technologies.
The demonstrated device operates at room temperature and telecom wavelength, is amenable to electrical pumping \cite{Boitier14} thanks to the direct bandgap of AlGaAs, and has a high potential of integration within photonic circuits \cite{Dietrich16}: the monolithical integration with on-chip beamsplitters has been demonstrated \cite{Belhassen18}, and the integration of electro-optic phase shifters \cite{Wang14,Dietrich16} for further manipulation of the state and superconducting nanowires to achieve on-chip detection \cite{Schwartz18} can be envisaged.
The used transverse pump configuration circumvents the usual issue of pump filtering and allows a direct spatial separation of the photons of each pair,  facilitating their use in protocols.
In particular, these results could be harnessed to study the effect of exchange statistics in various quantum simulation problems \cite{Crespi13,Matthews13,Crespi15} with a chip-integrated platform, and for communication and computation protocols making use of antisymmetric high-dimensional quantum states \cite{Jex03,Goyal14}.
Other non-Gaussian high-dimensional photonic states such as time-frequency Schr\"odinger cat or compass states could also be realized in the used device by a further engineering of the pump beam \cite{Boucher15}. In addition, direct generation of polarization entanglement has already been demonstrated with this source design \cite{Orieux13} and similar chip-integrated structures \cite{Horn13}, opening the perspective to combine such discrete-variable entanglement with the continuous-variable-like entanglement demonstrated here in the time-frequency degrees of freedom of the photon pairs.

\section*{Funding}

We acknowledge support from Agence Nationale de la Recherche (project SEMIQUANTROOM), Région Ile-de-France in the framework of the C’Nano DIM NanoK (project SPATIAL), European Union’s Horizon 2020 research and innovation programme under the Marie Skłodowska-Curie grant agreement No 665850, Labex SEAM (Science and Engineering for Advanced Materials and devices, ANR-10-LABX-0096), IdEx Université de Paris (ANR-18-IDEX-0001), and the French RENATECH network.

\section*{Acknowledgment}

The authors thank M. Apfel and F. Bouchard for technical support. 

\bigskip

\bibliography{D:/Travail/MPQ/Biblio}

\clearpage
\onecolumngrid
\begin{center}
	\textbf{\large Supplemental Material -- Engineering two-photon wavefunction \\ and exchange statistics in a semiconductor chip}
\end{center}

\setcounter{equation}{0}
\setcounter{figure}{0}

\renewcommand{\theequation}{S\arabic{equation}}
\renewcommand{\thefigure}{S\arabic{figure}}

\section{Methods}

\paragraph{Sample} The sample was grown by molecular beam epitaxy on a (100) GaAs substrate. The epitaxial structure consists
in a 36-period asymmetrical Al$_{0.35}$Ga$_{0.65}$As/Al$_{0.90}$Ga$_{0.10}$As
distributed Bragg reflector,
a 4.5-period Al$_{0.80}$Ga$_{0.20}$As/Al$_{0.25}$Ga$_{0.75}$As quasi-phasematching waveguide core and 14-period asymmetrical
Al$_{0.35}$Ga$_{0.65}$As/Al$_{0.90}$Ga$_{0.10}$As distributed Bragg reflector. The planar structure is then chemically etched to obtain a ridge of width 6 $\mu$m and height 7 $\mu$m. The ridge is then cleaved (leading to the formation of facets) to obtain a device length of 2 mm.

The refractive index contrast between the AlGaAs compound and the air induces a finite reflectivity of the facets of the waveguide. Finite-Difference Frequency-Domain (FDFD) numerical simulations allow to estimate the modal
reflectivities R$_{TE}$ = 26.7\% and R$_{TM}$ = 24.7\% for the TE and TM fundamental modes, respectively, at a wavelength of 1550 nm.

\paragraph{Source brightness}

The measured coincidence count rate is 50 Hz in the HOM experiment (see Fig. 4b) and the corresponding single count rate is 35 kHz. From this we deduce the internal pair production rate $\simeq12$ MHz. Taking into account losses in the sample ($\alpha \simeq 0.5$ cm$^{-1}$ for the down-converted modes) and facet transmission, the pair production rate at the chip output is $\simeq 10$ MHz, for a time-averaged pump power on the sample of $\simeq 50$ mW. The source brightness is thus $\simeq 200$ kHz/mW, which is for comparison about one order of magnitude higher than for the FWM glass microrings used in Ref. \cite{Kues17}.

\begin{figure}[h]
	\centering
	\includegraphics[width=0.45\columnwidth]{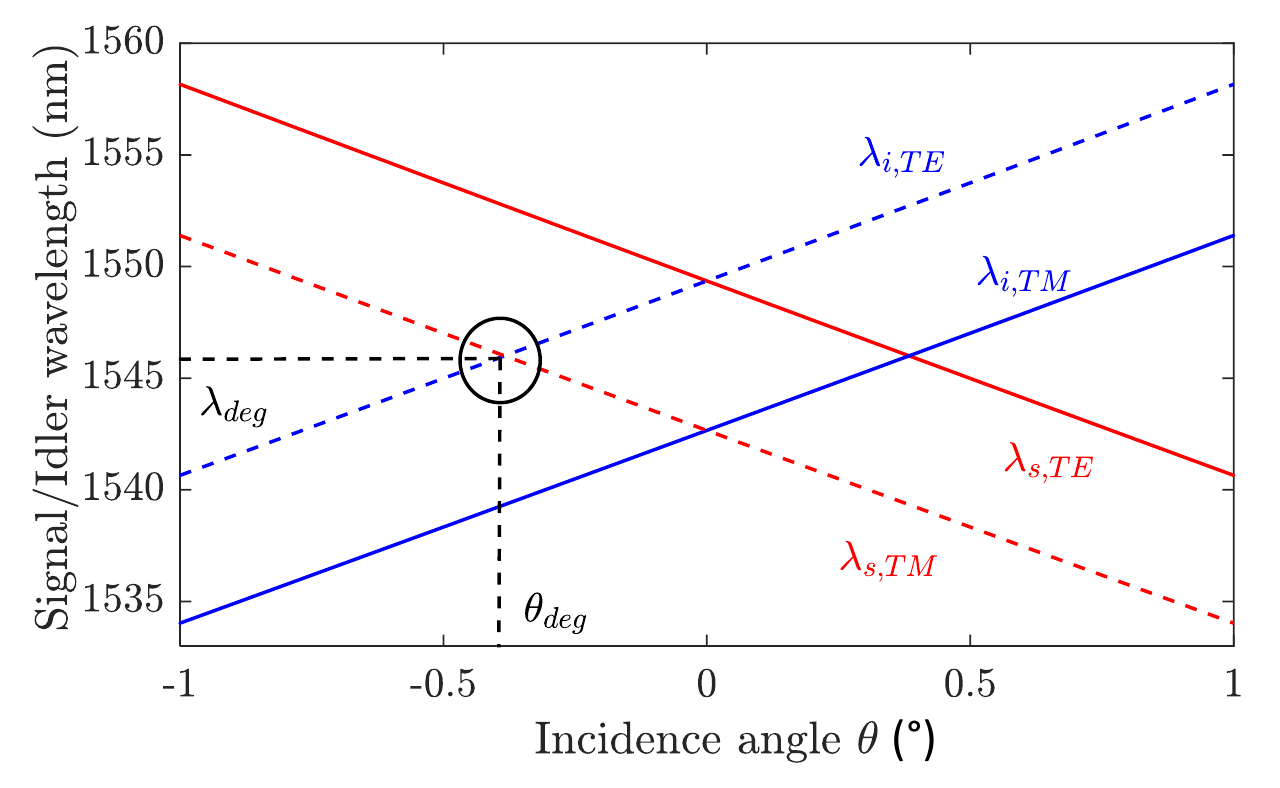}
	\caption{
		Angular tuning curve of the signal/idler photons generated by PDC for the source used in the experiments and a pump wavelength $\lambda_p=773$ nm. 
	}
	\label{FigTuningCurve}
\end{figure}

\paragraph{Spectral tuning of the source} We show in Fig. \ref{FigTuningCurve} the simulation for our source, of the wavelength of signal (red lines) and idler (blue lines) photons as a function of the pump beam incidence angle $\theta$, for the pump wavelength $\lambda_p=773$ nm used in the experiments.
Plain and dotted lines correspond to both possible type-II PDC interactions occurring in the device. In the experiment we consider the one (dotted lines) that generates a TM-polarized signal photon (propagating along $z>0$, see Fig. 1a) and a TE-polarized idler photon (propagating along $z<0$); the black circle indicates the degeneracy condition obtained by pumping the source with an incidence angle $\theta_{\rm deg}$.

\paragraph{Simulation of the JSA}

The JSA simulations in Fig. 2 and 3 are done numerically using a Matlab script. We consider the epitaxial structure of the source to calculate the effective indices of the modes in a 1D model adapted from Ref. \cite{Chilwell84}, using the AlGaAs material indices parametrization of Gehrsitz et al. \cite{Gehrsitz00}. We then calculate the phase-matching and energy conservation terms of the joint spectral amplitude (JSA) of the photon pairs using the formulas recalled in article. The JSA is later used to calculate the HOM probability and the Wigner function shown in Fig. 4.  
In the simulations the following approximations are made: we consider a low pump regime, so that we consider an undepleted pump and neglect any multi-photon component of the emitted state. We take into account frequency dispersion of the refractive indices but we neglect the group velocity dispersion (which is justified by the narrow spectral range of the generated photon pairs) \cite{Boucher15}.

\section{Theory}

\subsection{Control of the biphoton exchange statistics}

We here provide additional theoretical details showing how the manipulation of the biphoton spectral wavefunction can be used to simulate particles with different quantum statistics. 

\bigskip

We start by describing the algebra for creation and annihilation operators. The creation (resp. annihilation) operator for a particle in the frequency mode $\omega$ is defined as:
\begin{equation}
\hat{a}^{\dagger}(\omega)\ket{0}=\ket{\omega} \ , \ \hat{a}(\omega)\ket{\omega'}=\delta(\omega-\omega')\ket{0},
\end{equation}
where $\ket{0}$ denotes the vacuum. Their action is to increase (resp. decrease) the number of particles in mode $\omega$. These operators satisfy commutation relations that depend on the statistics of the quantum particles (controlled by the parameter $\Delta\varphi$):
\begin{align}
\hat{a}^{\dagger}(\omega)\hat{a}^{\dagger}(\omega')-e^{i\Delta\varphi}\hat{a}^{\dagger}(\omega')\hat{a}^{\dagger}(\omega)=0,\\
\hat{a}(\omega)\hat{a}(\omega')-e^{i\Delta\varphi}\hat{a}(\omega')\hat{a}(\omega)=0,\\
\hat{a}(\omega)\hat{a}^{\dagger}(\omega')-e^{-i\Delta\varphi}\hat{a}^{\dagger}(\omega')\hat{a}(\omega)=\delta(\omega-\omega')\mathbb{I}.
\end{align}
where $\mathbb{I}$ is the identity operator.
The case $\Delta\varphi=0$ corresponds to bosons, $\Delta\varphi=\pi$ to fermions, and intermediate values of $\Delta\varphi$ correspond to anyons \cite{Matthews13}.

\bigskip

The wavefunction describing two quantum particles in continuous frequency modes is:
\begin{equation}\label{JSA}
\ket{\psi}= \iint d \omega_s d \omega_i \, {\rm JSA} (\omega_s,\omega_i)\hat{a}^\dagger_{s}(\omega_s) \hat{a}^\dagger_{i} (\omega_i)\ket{0,0}_{s,i}
\end{equation}
where $\text{JSA}$ is the joint spectral amplitude, and the creation operators obey the above defined commutation relations. Depending on the symmetry properties of the JSA, we can simulate different quantum statistics as will be shown in Section 1.B.

\bigskip

As described in the main text, in the experiment we can control the symmetry of the biphoton states by pumping the semiconductor structure with a pump beam profile of the form $\mathcal{A}_p (z)=F(z) e^{-z^2/w^2}e^{ik z}$, where $F(z)=1$ for $z<0$ and $F(z)=e^{i \Delta \varphi}$ for $z>0$ (see Fig. 3a). The phase-matching function (see Eq. (2) of the main text) then reads:
\begin{equation}
\phi_{\rm PM}(\omega_s,\omega_i)=\int_{-L/2}^{0} dz \,e^{-z^2/w^2} e^{-i (\omega_s-\omega_i)z/v_{\rm g}}e^{i(k-k_{\text{deg}})z}+e^{i\Delta\varphi}\int_0^{L/2} dz \,e^{-z^2/w^2} e^{-i (\omega_s-\omega_i)z/v_{\rm g}}e^{i(k-k_{\text{deg}})z}
\end{equation}
When pumping at the degeneracy angle, $k=k_{\text{deg}}$, and a change of variable $(z\rightarrow -z)$ in the first integral leads to Eq. (3) of the main text:
\begin{equation}
\phi_{\rm PM}(\omega_s,\omega_i)=f(\omega_s,\omega_i)+e^{i \Delta \varphi} f(\omega_i,\omega_s)
\end{equation}
with $f(\omega_s,\omega_i)=\int_0^{L/2} dz \,e^{-z^2/w^2} e^{i (\omega_s-\omega_i)z/v_{\rm g}}$. 
For $\Delta \varphi=0 \;(\pi)$ the phase-matching function, and thus the JSA is symmetric (antisymmetric) under particle exchange, leading to a bosonic (fermionic) quantum statistics (see Section 1.B. below).

\bigskip

Note that a more explicit expression of the phase-matching function can be obtained in the limit $L \gg w$:
\begin{equation}
\phi_{\rm PM}(\omega_s,\omega_i)=w\left[e^{i\Delta\varphi} \text{fadf}\left(\frac{\omega_{-}w}{v_{g}}\right)-\text{fadf}\left(\frac{\omega_{-}w}{v_{g}}\right)+2e^{-(\omega_{-}w/v_{g})^{2}/4}\right],
\end{equation}
where \text{fadf} is the Faddeeva function, $\text{fadf}(x)=e^{-x^{2}}\text{erfc}(-ix)$, and \text{erfc} is the complex error function.

\bigskip

Anyonic statistics, corresponding to intermerdiate values of $\Delta\varphi$ in the commutation relations (S2)-(S4), cannot be obtained with the simple phase-step profile considered above but could be obtained in the following way. We need the phase-matching function to obey the relationship $\phi_{\rm PM}(\omega_s,\omega_i)=e^{i\Delta\varphi}\phi_{\rm PM}(\omega_i,\omega_s)$, i.e., since for our SPDC process this function only depends on the frequency difference $\omega_{-}=\omega_{s}-\omega_{i}$:
\begin{equation}
\phi_{\rm PM}(\omega_{-})=e^{i\Delta\varphi}\phi_{\rm PM}(-\omega_{-})
\end{equation}
A suitable function would be $\phi_{\rm PM}(\omega_{-})=\omega_{-}^{\alpha}\text{exp}(-\omega_{-}^{2}/\beta)$, with $\alpha$ and $\beta$ real; in practice we can choose $\alpha\in [0,2]$. Since the phase-matching function reads $\phi_{\rm PM}(\omega_{-})=\int_{-L/2}^{L/2} A(z)e^{-i\omega_{-}z/v_{g}} \text{d}z$, with $A(z)=\mathcal{A}_p (z) e^{-i k_{\rm deg}z}$, the needed pump profile can be obtained by inverse Fourier transform:
\begin{equation}
\mathcal{A}_p (z) \propto e^{i k_{\rm deg}z} \int \text{d}\omega_{-} e^{i\omega_{-}z/v_{g}} \phi_{\rm PM}(\omega_{-})
\end{equation}
which is valid if the pump profile if narrower than the length of the chip. The integral can be performed numerically to find the pump phase profile that will produce biphotons with anyonic statistics of phase $\Delta\varphi$. This pump profile can then be implemented with the SLM.

\subsection{Hong-Ou-Mandel coincidence probability and Chronocyclic Wigner function}

We here provide a brief demonstration that the HOM experiment allows to read the symmetry of the biphoton spectral wavefunction, leading to bunching when the JSA is symmetric and antibunching when the JSA is antisymmetric. We then relate the HOM coincidence probability to the chronocyclic Wigner function of the biphoton state, which is discussed at the end of the main text.

\bigskip

Starting from the biphoton state of Eq. (S5), a time delay $\tau$ is introduced in the HOM interferometer; after the beamsplitter, and considering only the part of the wavefunction that will give rise to coincidence events, the state can be written as:
\begin{equation}
\ket{\psi}=\frac{1}{2}\iint \text{d}\omega_{s} \text{d}\omega_{i}  \left(\text{JSA}(\omega_{s},\omega_{i})e^{i\omega_{s}\tau}-\text{JSA}(\omega_{i},\omega_{s})e^{i\omega_{i}\tau}\right)\hat{a}^{\dagger}(\omega_{s})\hat{b}^{\dagger}(\omega_{i})\ket{0,0}
\end{equation}
where we called a (b) the upper (lower) spatial port.
Considering that the detectors have a flat frequency response, the coincidence probability is $P(\tau)=\iint \text{d}\omega_{s}\text{d}\omega_{i} \abs{\bra{\omega_{s},\omega_{i}}\ket{\psi}}^{2}$, leading to:
\begin{equation}
P(\tau)=\frac{1}{2}\left(1-\text{Re} \iint \text{d}\omega_{s}\text{d}\omega_{i} \text{JSA}(\omega_{s},\omega_{i})\text{JSA}^{*}(\omega_{i},\omega_{s})e^{i\omega_{-}\tau}\right)
\end{equation}
where $\text{Re}$ denotes the real part.

If the JSA is symmetric under the exchange of the photons, then $ P(\tau)=\frac{1}{2}\left(1-\text{Re}\iint \text{d}\omega_{s}\text{d}\omega_{i} \text{JSI}(\omega_{s},\omega_{i})e^{i\omega_{-}\tau}\right)$. At zero delay we obtain $P(\tau=0)=0$ since the wavefunction is normalized: the photons bunch, which is the signature of bosonic statistics. If the JSA is antisymmetric under particle exchange, 
$ P(\tau)=\frac{1}{2}\left(1+\text{Re}\iint \text{d}\omega_{s}\text{d}\omega_{i} \text{JSI}(\omega_{s},\omega_{i})e^{i\omega_{-}\tau}\right)$,
and $P(\tau=0)=1$: the photons antibunch, corresponding to fermionic statistics. 

\bigskip

An equivalent way of deriving this result is to re-write the wavefunction (S11) in the case of fermions, for which
$\hat{a}^{\dagger}(\omega_{s})\hat{b}^{\dagger}(\omega_{i})=-\hat{b}^{\dagger}(\omega_{i})\hat{a}^{\dagger}(\omega_{s})$; the state $\ket{\psi}$ then reads:
\begin{equation}
\ket{\psi}=\frac{1}{2}\iint \text{d}\omega_{s} \text{d}\omega_{i}  \left(\text{JSA}(\omega_{s},\omega_{i})e^{i\omega_{s}\tau}+\text{JSA}(\omega_{i},\omega_{s})e^{i\omega_{i}\tau}\right)\hat{a}^{\dagger}(\omega_{s})\hat{b}^{\dagger}(\omega_{i})\ket{0,0}
\end{equation}
and the coincidence probability becomes:
\begin{equation}
P(\tau)=\frac{1}{2}\left(1+\text{Re} \iint \text{d}\omega_{s}\text{d}\omega_{i} \text{JSA}(\omega_{s},\omega_{i})\text{JSA}^{*}(\omega_{i},\omega_{s})e^{i\omega_{-}\tau}\right)
\end{equation}
Hence, in the case of fermions with a symmetric joint spectrum, we recover the behavior of bosons with an antisymmetric joint spectrum, i.e. $P(\tau=0)=1$.

\bigskip

In a previous work \cite{Douce13}, it was shown that the HOM coincidence probability can be related to the chronocyclic Wigner function of the biphoton state. Following the definition given in the main text, the phase-matching part of the Wigner function reads $W_{-}(\omega_{-},t_-)=\int \text{d}\omega'_- \, \phi_{\rm PM}(\omega_{-}+\omega'_-) \, \phi_{\rm PM}^{*}(\omega_{-}-\omega'_-) \, e^{i2\omega'_- t_-}$, where $\omega_{-}=\omega_{s}-\omega_{i}$ and $t_{-}=(t_{s}-t_{i})/2$. The coincidence probability in the HOM experiment is thus determined by the cut of the $W_{-}$ function along $\omega_-=0$:
\begin{equation}
P(\tau)=\frac{1}{2}\left(1-W_{-}(0,\tau)\right).
\end{equation}

\section{Visibility of the Hong-Ou-Mandel experiments}

We performed numerical simulations to understand the factors limiting the visibility of the HOM experiments reported in Fig. 4b and 4f of the article, obtained when pumping the source with a flat and $\pi$-phase step profile, respectively.

A first factor is the pump incidence angle $\theta$ (see Fig. 1a), which determines the wavelength of signal and idler photons. Spectral degeneracy occurs when pumping with the angle $\theta_{\rm deg}$ ; in the experiment, the spectral degeneracy is checked using the fiber spectrograph technique (see Fig. 1c), and we estimate a typical error of 
50 pm (wavelength difference between signal and idler) for this procedure, corresponding to $\sim 10 \%$ of the joint spectrum FWHM.
Taking into account this factor in the simulation, we estimate a visibility decrease of $7 \%$, for both pumping conditions (flat and $\pi$-step phase profile).
In addition, even when the signal and idler central frequencies are equal, their spectral overlap is not perfect due to the modal birefringence of the source, which leads to a slight displacement (15 pm in wavelength) between the Fabry-Perot resonances of the idler and signal photons. This leads to an additional visibility decrease of $1.5 \%$.

\begin{figure}[h]
	\centering
	\includegraphics[width=0.7\columnwidth]{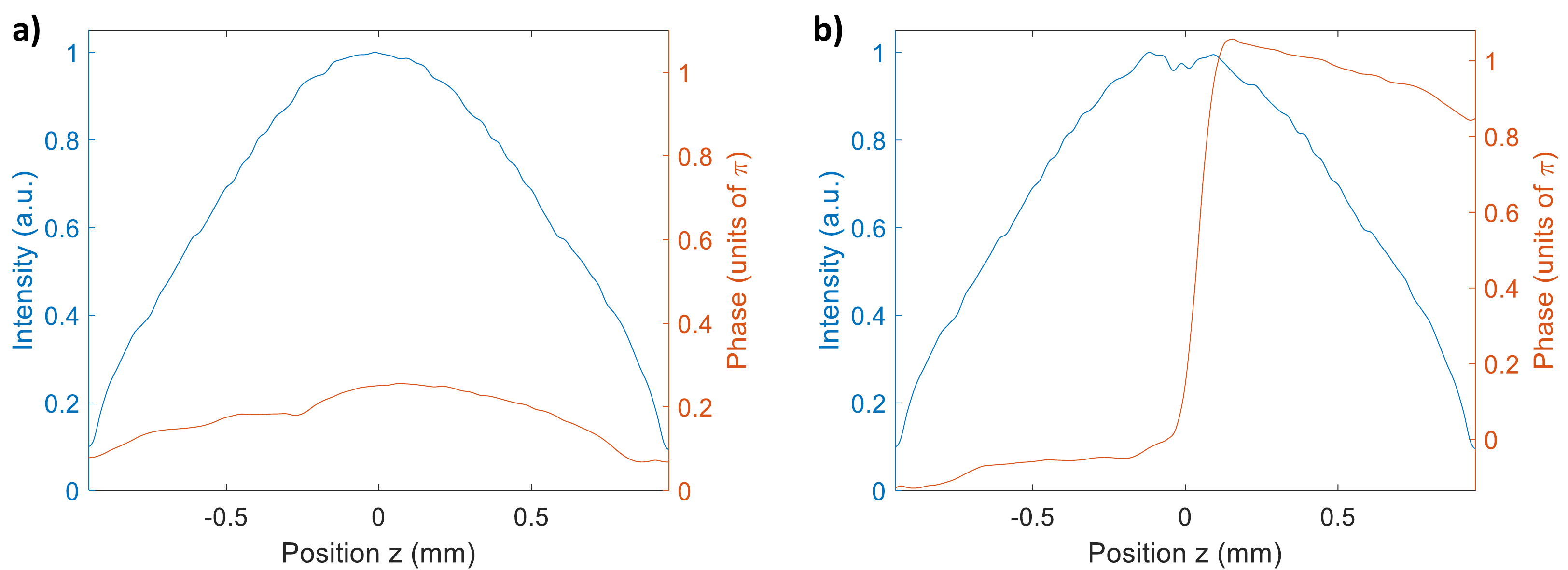}
	\caption{
		Measured phase (red) and intensity (blue) profiles of the pump beam used to produce the (a) symmetric and (b) antisymmetric biphoton frequency states.
	}
	\label{FigSM_spot}
\end{figure}

A second series of factors is related to the spatial properties of the pump beam. To evaluate the effect of the imperfections of the pump spot, we simulated the result of the HOM experiment by inputting the pump intensity and phase profile experimentally measured using a wavefront analyzer, as shown in Fig. \ref{FigSM_spot}. These experimental imperfections lead to a 8 \% (0.5 \%) visibility drop in the case of the $\pi$-step (flat) pump phase profile. Finally, the imprecision in the longitudinal centering of the pump spot on top of the waveguide, which is about 200 $\mu$m in our setup, leads to an additional 2 \% visibility drop for both phase pumping conditions.

By taking into account simultaneously all above sources of experimental imperfections, we simulate HOM visibilities of 
$82 \%$ in the case of the $\pi$-step profile, and $90 \%$ in the case of the flat phase profile, which are close to the experimental results ($77 \%$ and $88 \%$ respectively). The remaining $2-5 \%$ of visibility drop could be due to a slight polarization distinguishability between the two photons, which was not taken into account in the simulations.

This analysis shows that, in future experiments, a higher spectral resolution to adjust the degeneracy condition (which could be achieved, in the fiber spectrograph technique, using e.g. longer spools of DCF or superconducting detectors that have a shorter jitter) could allow up to a 7 \% visibility gain, while implementing a feedback loop on the SLM to correct for pump imperfections could allow an additional 8 \% visibility gain in the $\pi$-step pumping condition.

\section{Production of more multimode states}

Quantum states with higher Schmidt numbers (i.e. more time-frequency modes) could be obtained with the same source, along different pathways. 

Let us first consider the situation of a flat phase spatial profile as discussed in Fig. 2. A JSA with stronger anticorrelation could be obtained by pumping with a smaller pump waist (e.g. 0.1 mm could be easily achieved) and by using longer pump pulses (e.g. 10 ps using the same laser, or 1 ns using a nanosecond-pulsed laser). In these two cases the estimated Schmidt numbers are:
\begin{center}
	\begin{tabular}{|c|c|c|}
		\hline 
		Waist 0.1 mm	& Pulse duration 10 ps  & $K=4.5$ \\ 
		\hline 
		Waist 0.1 mm	& Pulse duration 1 ns & $K=440$ \\ 
		\hline 
	\end{tabular} 
\end{center}
All intermediate values of $K$ can be achieved by choosing a pulse duration between these two limits.

On the contrary, a JSA with stronger positive correlation could be obtained by pumping with a bigger pump waist (e.g. 2 mm could be easily achieved) and by using shorter pump pulses. In the latter case, the finesse of the vertical cavity (that confines the pump beam) currently limits the minimum intracavity pulse time to about 2 ps. However, fabricating a sample with a lower finesse for the vertical cavity (with less periods for the Bragg reflectors) would allow to use e.g. a pulse duration of 0.5 ps. In these two cases the estimated Schmidt numbers are:
\begin{center}
	\begin{tabular}{|c|c|c|}
		\hline 
		Waist 2 mm	& Pulse duration 2 ps  & $K=3.0$ \\ 
		\hline 
		Waist 2 mm	& Pulse duration 0.5 ps & $K=11$ \\ 
		\hline 
	\end{tabular} 
\end{center}

Another possibility is to play with the spatial phase of the pump beam. Fig. 3k of the article already shows e.g. that K can be increased from 1.4 to 2.2 by imposing a $\pi$ phase step at the center of the pump beam. On the other hand, applying a quadratic spatial phase (either using the SLM, or simply by focusing the pump beam on the sample surface) can efficiently increase the Schmidt number. For instance, starting from the uncorrelated JSA of Fig. 2c (waist 0.6 mm, pulse duration 6 ps), using a pump spot with an (experimentally feasible) curvature radius of 10 cm would yield $K\simeq7$.

\end{document}